\documentclass[12pt,tightenlines,eqsecnum,floats,showpacs,nofootinbib,amsmath,amssymb,aps,prd]{revtex4}

\usepackage{amsmath,amssymb,amsfonts,amsthm,amscd}
\usepackage{graphicx}
\usepackage{enumerate} 
\usepackage{colordvi} 
\usepackage{units}
\usepackage{epsfig}
\usepackage{color}
\usepackage[bookmarksopen]{hyperref}

\hypersetup{
pdfstartview = {FitH},
}

\newcommand{\bra}[1]{\left\langle #1\right|}
\newcommand{\ket}[1]{\left|#1\right\rangle}

\newcommand{\bk}[2]{\left\langle  #1|#2\right\rangle}

\newcommand{\abs}[1]{\left|#1\right|}

\newcommand{\norm}[1]{\left\|#1\right\|}

\DeclareMathOperator{\Tr}{Tr}
\DeclareMathOperator{\R}{Re}
\DeclareMathOperator{\I}{Im}
\DeclareMathOperator{\sl2c}{SL(2,\mathbb{C})}
\DeclareMathOperator{\su2}{SU(2)}
\DeclareMathOperator{\diag}{diag}

\DeclareMathOperator{\ach}{arcosh}

\begin{document}

\title{Anisotropic Spinfoam Cosmology}
\author{Julian Rennert$^{1,2}$}
\email{Rennert@stud.uni-heidelberg.de}
\author{David Sloan$^3$}
\email{djs228@hermes.cam.ac.uk}
\affiliation{$^1$ Centre de Physique Th\'{e}orique 1, CNRS-Luminy, Case 907, F-13288 Marseille\\
$^2$ Institut f\"{u}r Theoretische Physik, Universit\"{a}t Heidelberg, Philosophenweg 16, D-69120 Heidelberg\\
$^3$ DAMTP, Center for Mathematical Sciences, Cambridge University, Cambridge CB3 0WA, UK}

\begin{abstract}
The dynamics of a homogeneous, anisotropic universe are investigated within the context of spinfoam cosmology. Transition amplitudes are calculated for a graph consisting of a single node and three links - the `Daisy graph' - probing the behaviour a classical Bianchi I spacetime. It is shown further how the use of such single node graphs gives rise to a simplification of states such that all orders in the spin expansion can be calculated, indicating that it is the vertex expansion that contains information about quantum dynamics.
\end{abstract}


\maketitle

\newpage

\section{Introduction}

Spinfoam methods \cite{Zakopane} represent a novel and interesting quantization of gravity. These comprise the covariant formulation of loop quantum gravity (LQG) \cite{Abhay,rovelli}. In a previous paper \cite{ourpaper} we set the stage for the investigation of an anisotropic homogeneous spacetime within the spinfoam cosmology approach. This was done by proposing a simple graph, made from one node and three closed links which is used to probe anisotropic geometries. We showed that in the isotropic case this graph is sufficient to reproduce known results in spinfoam cosmology \cite{towards}, i.e. to show that the geometry can be assumed to be peaked on a flat, static spacetime.

The spinfoam approach to quantum gravity is a `bottom up' construction - one does not take a gravitational theory and directly quantize it but rather begins from a fundamental quantum theory and derives dynamics \cite{newlook}. It is therefore not obvious, and certainly non-trivial, that the theory results in agreement with general relativity (GR) in the classical limit. Thus it is important to test these ideas within simple contexts such as cosmological models.  The simplest such models are those which describe the homogeneous, isotropic Robertson-Walker geometries, whose dynamics form the Friedmann-Lem\^aitre solutions in GR. Within the spinfoam context, these have been examined in \cite{towards} and the resulting dynamics shown to agree in the classical limit. However, for FLRW models evolution is directly tied to the presence of matter - vacuum solutions are necessarily static, and thus not suited testing the agreement of dynamics. Although there are proposals for the coupling of fermions and Yang-Mills fields to spinfoam models \cite{matter1, matter2}, there is to date no detailed analysis of their asymptotic or semiclassical limit. Furthermore, it may be possible to include an effective matter coupling in a similar fashion as the cosmological constant \cite{cosmoconstant}, but this has not been investigated in a cosmological setting either, to the best of our knowledge. Thus, we must extend these models to obtain evolving solutions. Within the isotropic case, the inclusion of a cosmological constant has been examined and shown to give good agreement with the classical behaviour \cite{cosmoconstant}. In this paper we take a separate approach: by relaxing the isotropy of our geometry we access a dynamical vacuum and we will show that the graph used in \cite{ourpaper} is sufficient to derive vacuum Bianchi I geometries from the full quantum theory. This model proves an ideal test of spinfoam cosmology - it is simple enough to be tractable, but not so simple as to be static and requires no extra assumptions to be made beyond those used in establishing the spinfoam formalism. Furthermore, the classical solution is singular, a property which is hoped to be cured by quantum gravity effects. The resolution of classical singularities is a key feature of the canonical approach to LQG in the cosmological setting - loop quantum cosmology (LQC) \cite{APS}. Strong singularities have been shown to be cured by LQC in the $k=0$ \cite{bojo2,ParamLQCNeverSingular}, and $k=\pm 1$ \cite{ParamFrancesca} FLRW models. Recently this has been extended to include Bianchi I spacetimes \cite{ParamB1Singular}. Much of the potential for observable phenomena are based upon this resolution \cite{Measure,AbhayWillIvan}. Within the spinfoam framework we found hints for a possible singularity resolution in \cite{ourpaper} by interpreting the suppression of the amplitude for small spins as the avoidance of the classical singularity and in \cite{CarloFrancescaRecent} one finds in a much more general setting why singularities might be avoided in the new spinfoam models.

Bianchi models have been the subject of much work in LQC. Within the quantum framework \cite{EdB1,EdB2,EdB9} and the semi-classical effective framework \cite{AlexB1,DahWei,Cailleteau,DahWei2,Maartens,Gupt,AlexB2,Gupt2,AlexB9} they have been found to be non-singular and reproduce GR at low curvature scales. Examining closely the nature of these models in the spinfoam context should allow for deeper insight into the links between the canonical and covariant formulations of LQG. 

The model which we use is based around the `Daisy graph', consisting of a single node which forms the target and source of three links. As such, this graph is dual to a cube whose opposing faces have been identified to form a three torus or a patch of a homogeneous lattice with periodic boundary conditions. The use of a single node simplifies calculations considerably since group elements arising at the source and target of each link are identical. It has been argued \cite{cubulations,MartinsCubes} that a cubulation of space-time is a natural setting for examining spinfoams, as all space-times admit a decomposition into tesseracts, and thus initial and final hypersurfaces can be decomposed into cubes.  

This paper is laid out as follows: In section \ref{GenTheory} we recap the general theoretical foundations of the spinfoam formulation of LQG and outline results which are important to the model under consideration. Section \ref{Classical} gives the classical description of dynamics of a Bianchi I model in GR. In section \ref{OurModel} we show that the transition amplitude for the Daisy graph is in agreement with the GR result. Finally we end in section \ref{discussion} with a discussion of the results and suggestions for extensions to this model.

\section{General theory}
\label{GenTheory}

In this section we shall accumulate some of the theoretical foundations from LQG and spinfoams which will be used in the course of this work. In particular, we will use the holomorphic \cite{holo} / complexifier \cite{gcs1,gcs2} coherent states, the EPRL-FK/KKL vertex amplitude \cite{spinfoam1,spinfoam2,spinfoam3,spinfoam4,spinfoam5} and show how one can normalize the coherent states following \cite{gcs1,gcs2}. We will not review how the spinfoam model is constructed but rather use the simple heuristic picture of a spinfoam as arising from the dynamics of the canonical spin network states. For our general motivation and arguments for certain approximations to be made, cf. \cite{ourpaper}. Throughout this paper we denote $\su2$ elements by $h$ or $u$ and $\sl2c$ elements by $g$, $G$ or $H$. The polar decomposition of $\sl2c$ is written as $g = M u$, with $u \in \su2$ and $M$ hermitian.

\subsection{LQG and spinfoam vertex amplitude}

We consider the canonical Hilbert space $\mathcal{H}_{\Gamma} = L_2(SU(2)^L/SU(2)^N)$, \cite{rovelli,thiemann,Zakopane}, which gives a truncation of the full kinematical Hilbert space of LQG on a fixed graph $\Gamma$, (oriented, with $L$ links and $N$ nodes). The elements of this Hilbert space, the spin network functions, are quantum states of the gravitational field and a general function $\psi : \su2^L \rightarrow \mathbb{C}$ is projected onto its gauge invariant part via
\begin{equation}
\Psi(h_l) = P_{\su2}\psi(h_l) = \int_{\su2}{du_l \, \psi(u_{s(l)} h_l u^{-1}_{t(l)})}\,.
\label{eq:1}
\end{equation}

Gauge transformations only act on the nodes of a graph and we denote the source and the target node of the link $l$ as $s(l)$ and $t(l)$ respectively. Each link of the graph is equipped with two variables: The holonomy of the Ashtekar connection
\begin{equation}
h_l = h_l[A] = \mathcal{P} \exp \left(\int_l{A}\right) \qquad \text{with} \qquad A^i_a = \Gamma^i_a + \gamma K^i_a
\label{eq:2}
\end{equation}

($\mathcal{P}$ denotes the path ordering symbol) and the flux variable
\begin{equation}
E(S_l) = \int_{S_l}{(\ast E)^j n^j}\,,
\label{eq:3}
\end{equation}
where $E(S_l)$ is the flux of the electric field $E=E^a_i \tau^i X_a$ through the surface $S_l$, which is dual to the link $l$. (The $\tau^i$ are $2i$ times the Pauli matrices and provide a basis for $\mathfrak{su}(2)$ and the $n^j$ are $\mathfrak{su}(2)$ valued smearing functions \cite{thiemann}.)

A spinfoam model now defines a notion of dynamics or evolution between states, which can be thought of as living on different spatial slices of a 3+1 - decomposition (for disconnected boundary). Thus, it tells us how to calculate transition amplitudes between spin network functions \cite{oeckl1,oeckl2}. Similar to the definition of $\su2$ spin networks on graphs the spinfoam is given by a 2 - Complex, build from vertices, edges and faces. In this heuristic picture, where a spinfoam results from the evolution of the spin network, the nodes of the graph become the edges and the links become the faces of the 2 - Complex. The dynamical information is incoded in the vertex amplitude and the full amplitude is given by the sum of different such spin network histories (or a refinement of the 2 - Complex \cite{refining}) which connect the same boundary spin networks. In this way the spinfoam formalism provides a discretized version for the path integral for quantum gravity.

The construction of the EPRL-FK/KKL spinfoam model requires an embedding of the $\su2$ spin network functions into $\sl2c$ spin network functions. This is done by using the (unitary) $Y_{\gamma}$ map
\begin{equation}
Y_{\gamma} : \mathcal{H}^{(j)} \rightarrow \mathcal{H}^{(\gamma j,j)}\,, \quad \ket{j,m} \mapsto \ket{(\gamma j, j); j, m}\,, \quad \mathcal{H}^{(p,k)} \cong \oplus_{j \geq k} \mathcal{H}^{(j)}\,,
\label{eq:4}
\end{equation}

which embeds the standard $\su2$ representation space $\mathcal{H}^{(j)}$ into to lowest weight of the decomposition of the $\sl2c$ representation space of the principal series, i.e. $\mathcal{H}^{(p,k)}$, with $p \in \mathbb{R}_{\geq 0}$ and $2k \in \mathbb{N}_{>0}$ such that $p=\gamma j$ and $k=j$. This then allows us to map functions on $\su2$ into functions on $\sl2c$ and group averaging again maps onto the gauge invariant states
\begin{equation}
\psi(h_l) \mapsto \Psi(g_l) = (P_{\sl2c} \circ Y_{\gamma} \psi)(g_l) = \int_{\sl2c}{dG'_n \, \psi(G_{s(l)} \, g_l \, G^{-1}_{t(l)})}\,,
\label{eq:5}
\end{equation}

where we have to neglect one integration over $\sl2c$, otherwise we would obtain a divergent factor as per \cite{reg}. When we write $\int_G{dG'_n}$ we mean $\int_{G^{N-1}}{dG_1...dG_{N-1}}$. The partition function is now defined as, \cite{Zakopane}
\begin{equation}
Z_{\mathcal{C}} = \int_{\su2}{dh_{vf} \, \prod_f{\delta(h_f)} \prod_v{A_v(h_{vf})}} \quad , \quad h_f = \prod_{v \subset f}{h_{vf}}
\label{eqnewnew:1}
\end{equation}

with the Lorentz-invariant vertex amplitude
\begin{equation}
A_v(h_l) = \int_{\sl2c}{dG'_n \, \prod_l{P(h_l,G_{t(l)} G^{-1}_{s(l)})}}
\label{eq:6}
\end{equation}

as a function of the boundary holonomies of the vertex. The Kernel $P$ can be given in the following form
\begin{equation}
P(h,g) = \sum_{2j \in \mathbb{N}_0}{(2j+1) \Tr(\mathcal{D}^{(j)}(h) Y^{\dagger}_{\gamma} \overline{\mathcal{D}^{(\gamma j,j)}(g)} Y_{\gamma})}\,,
\label{eq:7}
\end{equation}

where $\mathcal{D}^{(j)}(h)$ are the $\su2$ Wigner matrices and $\mathcal{D}^{(\gamma j,j)}(g)$ are the $\sl2c$ representation matrices. Now that one has a covariant definition of the vertex amplitude and the gauge invariant boundary states one can give a rigorous definition to the formal expression
\begin{equation}
W[ \, g^{(3)}_{\text{out}}, g^{(3)}_{\text{in}}] = \int^{g^{(3)}_{\text{out}}}_{g^{(3)}_{\text{in}}}{\mathcal{D}g^{(4)} \, e^{\frac{i}{\hbar} S[g^{(4)}]}}
\label{eq:8}
\end{equation}

in terms of a summation over spin network histories $\sigma$. For a state $\Psi$ on the boundary, i.e. a function of the boundary holonomies, and a spinfoam model $W$ the amplitude that corresponds to Eq.(\ref{eq:8}) is defined by \cite{holo}
\begin{equation}
\bk{W}{\Psi} = \int_{\su2}{dh_l \, W(h_l) \Psi(h_l)}\,,
\label{eq:9}
\end{equation}

where $W(h_l)$ is given by a sum over different spin network histories or 2-Complexes with a fixed boundary graph, an integration over bulk holonomies and a product of face amplitudes and the vertex amplitude given in Eq.(\ref{eq:6})
\begin{equation}
W(h_l) = \sum_{\sigma}{\int{dh^{\text{bulk}}_{vl} \prod_{f \subset \sigma}{\delta(h_f)} \prod_{v \subset \sigma}{A_v(h_{vl})} }}\,.
\label{eq:10}
\end{equation}

The argument of the delta function, i.e. $h_f$, is an oriented product of the holonomies on the links bounding the face $f$ and the $h_{vl}$ live on the links $l$, which are obtained as the intersection of a small 3-sphere around the vertex $v$ and the 2-Complex. Notice the correspondence of the integration over the holonomies in Eq.(\ref{eq:9}) together with the integration over the bulk holonomies in Eq.(\ref{eq:10}) to the measure in Eq.(\ref{eq:8}). One can think of this separation as a split into a time direction (integration along the bulk holonomies) and the spatial slice (integration along the $h_l[A]$).

In the next section we summarize briefly a few necessary facts about coherent states and their normalization before we consider the transition amplitude in the holomorphic representation and its normalization. As explained in \cite{gcs1,gcs2} a coherent state on a single link can be defined as 
\begin{equation}
\psi^t_g(h) = \sum_{2j \in \mathbb{N}_0} (2j+1) e^{- \frac{t}{2} j (j+1)}\Tr_j(g h^{-1})\,,
\label{eq:11}
\end{equation}

where $g \in \sl2c$ contains the information about the position in the classical phase space, where the state is peaked, i.e. $(A^i_a,E^a_i)$ and $t$ is the heat kernel time. A coherent state on the whole graph $\Gamma$ is then given by the tensor product of these single states, one for each link. If we introduce the polar decomposition of $\sl2c$, i.e. $g=Mu$, with $u \in \su2$, we have
\begin{equation}
\psi^t_g(h) = \psi^t_{M}(h u^{-1}) = \psi^t_{M u h^{-1}}(1) = \psi^t_{g h^{-1}}(1)\,,
\label{eq:12}
\end{equation}

which can be use, together with the following orthogonality relation
\begin{equation}
\int_{\su2}{dh \, \overline{\mathcal{D}^{(j)}(h)_{mn}} \, \mathcal{D}^{(j')}(h)_{op}} = \frac{1}{2j+1} \delta_{j j'} \delta_{m o} \delta_{n p}\,,
\label{eq:13}
\end{equation}

to show that the inner product between two gauge-variant coherent states is given by
\begin{equation}
\bk{\psi^t_{g}}{\psi^t_{g'}} = \psi^{2t}_{M M'}(h)\,,
\label{eq:14}
\end{equation}

where $g=Mu$, $g'=M'u'$ and $h=u^{-1}u'$. Thus, the norm of such a coherent state is given by
\begin{equation}
\norm{\psi^t_g}^2 = \psi^{2t}_{M^2}(1)
\label{eq:15}
\end{equation}

and the explicit formula is
\begin{equation}
\norm{\psi^t_g}^2 = \frac{4 \sqrt{\pi} \, e^{t/4}}{t^{3/2}} \frac{1}{\sqrt{y^2 - 1}} \sum^{\infty}_{n= - \infty}{ (\ach(y) - 2 \pi i n) \, e^{-\frac{(2 \pi n + i \ach(y))^2}{t}}}
\label{eq:16}
\end{equation}

with $y= \frac{1}{2} \Tr(g g^{\dagger}) = \frac{1}{2} \Tr(M^2)$. By writing $H = \exp \left(-i p_j \sigma^j / 2\right)$ and defining $p \equiv \sqrt{p_j p^j}$ one finds $y = \cosh(p)$. Furthermore, the character of any $g \in \sl2c$ can be calculated as follows: We write
\begin{equation}
g = \begin{pmatrix}a&b\\c&d\end{pmatrix} \quad \text{with} \quad ad - bc = 1 \quad , \quad a, b, c, d \in \mathbb{C}
\label{eq:17}
\end{equation}

and can give the explicit form of $g$ in the $j$-representation
\begin{equation}
\mathcal{D}^{(j)}(g)_{mn} = \sum_l{\frac{\sqrt{(j+m)! (j-m)! (j+n)! (j-n)!}}{(j-m-l)! (j+n-l)! (m-n+l)! l!}} \, a^{j+n-l} d^{j-m-l} b^{m-n+l} c^l\,.
\label{eq:18}
\end{equation}

The character is now given by\footnote{For diagonal $g$, i.e. $a=\lambda, b=c=0, d=\lambda^{-1}$, Eq.(\ref{eq:18}) reduces to $\mathcal{D}^{(j)}(g)_{mm} = a^{j+m} d^{j-m} = \lambda^{2m}$. Thus, one gets $\chi_j(g) = \sum_m{\lambda^{2m}} = (\lambda - \lambda^{-1})/(\lambda - \lambda^{-1}) \sum_m{\lambda^{2m}}$, $m \in \{-j,...,j\}$, which gives Eq.(\ref{eq:19}).}
\begin{equation}
\chi_j(g) = \Tr(\mathcal{D}^{(j)}(g)) = \frac{\lambda^{2j+1} - \lambda^{-(2j+1)}}{\lambda - \lambda^{-1}}\,,
\label{eq:19}
\end{equation}

which is an example of the Weyl character formula \cite{thiemann3}. For its derivation one uses the fact that the character is a class function (invariant under conjugation $g \mapsto h g h^{\dagger}$), thus $g$ can be assumed to be diagonal and from $\det(g) = \lambda_1 \lambda_2 = 1$ and $\Tr(g) = \lambda_1 + \lambda_2 = a + d$ one concludes
\begin{equation}
\lambda_1 = \lambda = x + \sqrt{x^2 - 1} \quad , \quad \lambda_2 = \lambda^{-1} = x - \sqrt{x^2 - 1}\,,
\label{eq:20}
\end{equation}
                
with $x = (a+d)/2$.

\subsection{Transition amplitude in the holomorphic representation and normalization}

We can now express the vertex amplitude and the transition amplitude in terms of the holomorphic coherent states as follows \cite{holo}. Take one coherent state per link, Eq.(\ref{eq:11}), tensor them together and group average over them. Such a gauge invariant coherent state $\Psi$, labelled by an $\sl2c$ element $H_l$ per each link, is given by
\begin{equation}
\Psi_{H_l}(h_l) = \int_{\su2}{du_l \prod_l{\sum_{2j \in \mathbb{N}_0}{(2j+1) \, e^{-\frac{t}{2} j (j+1)} \Tr_j(u_{s(l)} h_l u^{-1}_{t(l)} H^{-1}_l)}}}\,.
\label{eq:21}
\end{equation}

The vertex amplitude then takes the following form
\begin{equation}
A_v(H_l) = \int_{\sl2c}{dG'_n \, \prod_l{K_t(H_l,G_{t(l)} G^{-1}_{s(l)})}}\,,
\label{eq:22}
\end{equation}

where the Kernel is given by
\begin{equation}
K_t(H,G) = \sum_{2j \in \mathbb{N}_0}{(2j+1) \, e^{-\frac{t}{2} j (j+1)} \Tr(\mathcal{D}^{(j)}(H) Y^{\dagger}_{\gamma} \overline{\mathcal{D}^{(\gamma j,j)}(G)} Y_{\gamma})}
\label{eq:23}
\end{equation}

Using this representation and Eq.(\ref{eq:9}) the transition amplitude, which properly defines Eq.(\ref{eq:8}), becomes
\begin{align}
\bk{W}{\Psi} &= \int_{\su2}{dh_l \, W(h_l) \Psi_{H_l}(h_l)}\notag\\
&= \sum_{\sigma}{\int_{\su2}{dh_l \int{dh^{\text{bulk}}_{vl} \prod_{f \subset \sigma}{\delta(h_f)} \prod_{v \subset \sigma}{A_v(h_{vl})} }} \, \Psi_{H_l}(h_l)}\,.
\label{eq:24}
\end{align}

Now, if we want to calculate a physical transition amplitude we have to normalize it. In our model we will consider states which can be written as in - and out - states, i.e. $\Psi = \ket{\Psi_{\text{out}}} \otimes \ket{\Psi_{\text{in}}}$, which seems reasonable for the one-vertex spinfoam expansion and furthermore leads to the factorization of our amplitude\footnote{Note that for general boundary states it's not a priori clear that one can write a state in such a tensor form. The implicit assumption which allows us to write our state in such a form is the distinguishability of a past state and a future state, which is captured in our case by a disconnected boundary graph. Cf. \cite{mixedboundary}.}. We get
\begin{equation}
\bk{W}{\Psi} = \bra{W}(\ket{\Psi_{\text{out}}} \otimes \ket{\Psi_{\text{in}}}) = \bk{W}{\Psi_{\text{out}}} \bk{W}{\Psi_{\text{in}}}\,.
\label{eq:25}
\end{equation}

Hence, we find for the normalized transition amplitude
\begin{equation}
\mathcal{A}(H_l) = \frac{\bk{W}{\Psi}}{\bk{\Psi}{\Psi}} = \frac{\bk{W}{\Psi}}{\prod_l{\bk{\Psi_l}{\Psi_l}}} = \frac{\bk{W}{\Psi}}{\prod_l{\norm{\Psi_l}^2}}
\label{eq:26}
\end{equation}

where we use Eq.(\ref{eq:16}) for the norm of a coherent state on a single link. This way of normalizing the amplitude differs from the way the amplitude was normalized in \cite{towards,cosmoconstant} and \cite{ourpaper} in that we don't fix the heat kernel time and the $p=\I(z)$ of the denominator, cf. Eq.(\ref{eq:16}). This has the effect that certain terms (linear and quadratic in $\I(z)$) survive, which were supposed to cancel before. This way of normalizing corresponds to the procedure amplitudes are normalized in ordinary QFT. However, our main result of this paper is not influenced by the way this normalization is performed and we will also give reasons why this way seems to be the correct way of normalizing at the end of section \ref{amplitude}.

\section{Classical Bianchi I cosmology}
\label{Classical}

The Bianchi classification of 3-space describes homogeneous geometries \cite{Taub}. Of particular importance are those geometries which admit a Hamiltonian formulation - the class A geometries. This classification is in terms of 1-forms $\sigma_i$ which are related by 

\begin{equation}
 d \sigma_i = n_i \epsilon^{ijk} \sigma_j \wedge \sigma_k
\end{equation}

where the $n_i = 0, \pm 1$. It has been conjectured by Belinskii, Khalatnikov and Lifschitz \cite{BKL} that on approach to singularity all spacetimes become local, vacuum dominated and oscillatory. The BKL conjecture thus reduces the complexity of singularities in general to those of type VIII or IX. This conjecture has been the subject of many investigations, both analytic \cite{Andersson} and numerical \cite{Berger, Garfinkle}, and it is believed that a quantization of the resulting reduced dynamics may capture much of the character of full quantum gravity \cite{AHS, AHS2}. 

Although the conjecture results in cosmologies which are of type IX or VIII, these behave asymptotically like type I solutions for large periods \cite{Uggla}. Therefore it is likely that such models are both the simplest and the most relevant homogeneous cosmologies. 

\subsection{Metric structure}

The Bianchi I model is now given by $n_i=0$ and has the following line element 
\begin{equation}
ds^2 = - dt^2 + a^2_1(t) dx^2 + a^2_2(t) dy^2 + a^2_3(t) dz^2\,.
\label{eq:3.0}
\end{equation}

with directional scale factors $a_1(t), a_2(t), a_3(t)$. It corresponds to the diagonal metric $g_{\mu \nu}$ with inverse $g^{\mu \nu}$ and determinant $g$
\begin{equation}
g_{\mu \nu} = \diag(-1, a^2_1(t), a^2_2(t), a^2_3(t)) \quad , \quad g \equiv \det{g_{\mu \nu}} = - a^2_1(t) a^2_2(t) a^2_3(t)\,.
\label{eq:3.1}
\end{equation}

from which the 4-dim Ricci scalar $^{(4)} R$ is computed to be
\begin{equation}
^{(4)} R = \frac{2}{a_1 a_2 a_3} \left(a_1 \dot{a}_2 \dot{a}_3 + \dot{a}_1 a_2 \dot{a}_3 + \dot{a}_1 \dot{a}_2 a_3 + a_1 a_2 \ddot{a}_3 + a_1 \ddot{a}_2 a_3 + \ddot{a}_1 a_2 a_3\right)\,.
\label{eq:3.2}
\end{equation}

This leads to the following expression for the Einstein-Hilbert action
\begin{align}
S_{\text{EH}}\left[g_{\mu \nu}\right] &= \frac{1}{16 \pi G} \int_{\mathcal{M}}{^{(4)} R \sqrt{-g} \, d^4x}\notag\\
&= \frac{1}{8 \pi G} \int_{\mathcal{M}}{\left(a_1 \dot{a}_2 \dot{a}_3 + \dot{a}_1 a_2 \dot{a}_3 + \dot{a}_1 \dot{a}_2 a_3 + a_1 a_2 \ddot{a}_3 + a_1 \ddot{a}_2 a_3 + \ddot{a}_1 a_2 a_3\right) \, d^4x}\,.
\label{eq:3.3}
\end{align}

If we consider a universe with no matter and no cosmological constant, i.e. $(\Lambda = \rho = 0)$, the geometry is determined as a solution of the field equations $G_{\mu \nu} = 0$ and one finds that the components of the Einstein tensor for this model are given by
\begin{equation}
\begin{array}{l l}
      G_{00} = \frac{1}{a_1 a_2 a_3} \left(a_1 \dot{a}_2 \dot{a}_3 + \dot{a}_1 a_2 \dot{a}_3 + \dot{a}_1 \dot{a}_2 a_3\right) \, , \quad & G_{11} = - \frac{a^2_1}{a_2 a_3} \left(\dot{a}_2 \dot{a}_3 + \ddot{a}_2 a_3 + a_2 \ddot{a}_3\right) \, ,\\
			& \\
      G_{22} = - \frac{a^2_2}{a_1 a_3} \left(\dot{a}_1 \dot{a}_3 + a_1 \ddot{a}_3 + \ddot{a}_1 a_3\right) \, , \quad & G_{33} = - \frac{a^2_3}{a_1 a_2} \left(\dot{a}_1 \dot{a}_2 + a_1 \ddot{a}_2 + \ddot{a}_1 a_2\right) \, .\\
    \end{array}
\label{eq:3.4}
\end{equation}

Now, before we investigate possible solutions for this model let us have a look at its Hamiltonian structure and the boundary term
\begin{equation}
S_B \left[g_{\mu \nu}\right] = \frac{1}{8 \pi G} \oint_{\partial \mathcal{M}}{\epsilon K \sqrt{h} \, d^3y}\,,
\label{eq:3.5}
\end{equation}

where $\epsilon$ is $\pm 1$ and describes the direction of the normal vector of $\partial \mathcal{M}$. $h$ is the determinant of the induced 3-metric on $\partial \mathcal{M}$ and $K$ is the trace of the extrinsic curvature of $\partial \mathcal{M}$ \cite{poisson}. For the investigation of the Hamiltonian structure one uses a $3+1$ - split of the spacetime manifold and one can write the line element as
\begin{equation}
ds^2 = - N^2 dt^2 + h_{ab} (dy^a + N^a dt) (dy^b + N^b dt)
\label{eq:3.6}
\end{equation}

with the three metric $h_{ab}$, i.e. the induced metric on the spatial slice, the lapse function $N$ and the shift vector $N^a$. The Hamilton equations are obtained by variation of the Hamiltonian with respect to the independent variables $N$, $N^a$, $h_{ab}$ and $p^{ab}$ and variation with respect to lapse and shift give rise to the Hamiltonian constraint and the diffeomorphism constraint \cite{poisson,thiemann}
\begin{equation}
\mathcal{C} = \frac{1}{16 \pi G} \left( ^{(3)}R + K^2 - K^{ab} K_{ab}\right) \approx 0 \quad, \quad \mathcal{C}_a = \frac{1}{16 \pi G} \nabla_b \left(K^b_a - K \, \delta^b_a\right) \approx 0\,.
\label{eq:3.13}
\end{equation}

In order to find the full spacetime metric one picks initial data on a spatial slice, which has to satisfy the two constraints Eq.(\ref{eq:3.13}) and then one can use the Einstein equations in the Hamiltonian form to find their time evolution. Note that in our model $^{(3)}R = 0$ for all $a_1, a_2, a_3$, not just for solutions. Furthermore, the covariant derivative $\nabla_b$ in Eq.(\ref{eq:3.13}) reduces to a normal partial derivative in our model, because the spatial Christoffel symbols vanish. And since all the scale factors are just functions of time, not the coordinates, the diffeomorphism constraint is identically satisfied.

By comparison with the above line element (\ref{eq:3.0}), we realize that our metric already has the right form with $N=1$, $N^a = (0,0,0)$ and 
\begin{equation}
h_{ab} = \diag(a^2_1(t), a^2_2(t), a^2_3(t)) \quad , \quad h^{ab} = \diag(a^{-2}_1(t), a^{-2}_2(t), a^{-2}_3(t))\,.
\label{eq:3.7}
\end{equation}

Now, the extrinsic curvature of our spatial slice is given by $K_{ab} = \frac{1}{2} \mathcal{L}_{\partial_t} h_{ab} = \diag(a_1 \dot{a}_1, a_2 \dot{a}_2, a_3 \dot{a}_3)$. Furthermore, we need
\begin{equation}
K^{ab} = h^{ac} h^{bd} K_{cd} = \diag \left(\frac{\dot{a}_1}{a^3_1}, \frac{\dot{a}_2}{a^3_2}, \frac{\dot{a}_3}{a^3_3}\right)
\label{eq:3.8}
\end{equation}

from which we find the trace
\begin{equation}
K = h^{ab} K_{ab} = \frac{\dot{a}_1}{a_1} + \frac{\dot{a}_2}{a_2} + \frac{\dot{a}_3}{a_3}\,.
\label{eq:3.9}
\end{equation}

\subsection{Boundary term of the action}

With the above results we find for the boundary term Eq.(\ref{eq:3.5})
\begin{equation}
S_B \left[h_{ab}\right] = \frac{\epsilon}{8 \pi G} \oint_{\partial \mathcal{M}}{\left(\frac{\dot{a}_1}{a_1} + \frac{\dot{a}_2}{a_2} + \frac{\dot{a}_3}{a_3}\right) \, a_1 a_2 a_3 \, d^3y}\,.
\label{eq:3.11}
\end{equation}

Now we know that the boundary $\partial \mathcal{M}$ can be decomposed into two spacelike hypersurfaces $\Sigma_{t_1}$ and $\Sigma_{t_2}$ with $t_2 > t_1$ and a timelike boundary $\mathcal{B}$ such that we get for $S_G = S_{EH} + S_B$
\begin{equation}
(16 \pi G) S_G = \int_{\mathcal{M}}{^{(4)} R \sqrt{-g} \, d^4x} + 2 \int_{\Sigma_{t_2}}{K \sqrt{h} \, d^3y} - 2 \int_{\Sigma_{t_1}}{K \sqrt{h} \, d^3y} + 2 \int_{\mathcal{B}}{\mathcal{K} \sqrt{-\gamma} \, d^3z}\,,
\label{eq:3.12}
\end{equation}

where the minus sign in front of the $\Sigma_{t_1}$ contribution comes from the fact that the normals of $\partial \mathcal{M}$ should be outward pointing, for details cf. \cite{poisson}. For solutions of the field equations we have $^{(4)} R = 0$ and the last term prevents the action from diverging for non-compact spatial slices. Since we will be examining a spatial slice with three torus topology this term is zero in our model. If we now integrate Eq.(\ref{eq:3.3}) by parts we get
\begin{align}
&S_{\text{EH}}\left[g_{\mu \nu}\right] = \frac{1}{8 \pi G} \int_{\Sigma_t}{d^3y} \int^{t_2}_{t_1}{dt} \, \left(a_1 \frac{d}{dt}(a_2 \dot{a}_3) + a_2 \frac{d}{dt}(a_3 \dot{a}_1) + a_3 \frac{d}{dt}(a_1 \dot{a}_2)\right)\label{eq:3.22}\\[0.5\baselineskip]
&= \frac{1}{8 \pi G} \int_{\Sigma_t}{d^3y} \: \left[ \Big.\: a_1 a_2 \dot{a}_3 + a_1 \dot{a}_2 a_3 + \dot{a}_1 a_2 a_3 \: \Big|^{t_2}_{t_1} - \int^{t_2}_{t_1}{dt} \, \left( a_1 \dot{a}_2 \dot{a}_3 + \dot{a}_1 a_2 \dot{a}_3 + \dot{a}_1 \dot{a}_2 a_3 \right)\right]\notag
\end{align}

so that we get for the full gravitational action, Eq.(\ref{eq:3.12}), $\, (16 \pi G) S_G =$
\begin{equation}
- \, 2 \int_{\mathcal{M}}{\left( a_1 \dot{a}_2 \dot{a}_3 + \dot{a}_1 a_2 \dot{a}_3 + \dot{a}_1 \dot{a}_2 a_3 \right) \: d^4x \, + \, 4 \int_{\Sigma_{t_2}}{(a_1 a_2 \dot{a}_3 + a_1 \dot{a}_2 a_3 + \dot{a}_1 a_2 a_3) \, d^3y}}\notag
\end{equation}
\begin{equation}
- \, 4 \int_{\Sigma_{t_1}}{(a_1 a_2 \dot{a}_3 + a_1 \dot{a}_2 a_3 + \dot{a}_1 a_2 a_3) \, d^3y}\,.
\label{eq:3.23}
\end{equation}

The main result of this paper will be the derivation of this action from the EPRL-FK/KKL spinfoam model in a reduced cosmological setting, where the contribution of the bulk term vanishes for solutions due to the Einstein equation $\, G_{00} = 0 \,$ (Hamilton function).

\subsection{Ashtekar variables}
Based on the formula $\, ds^2 = -dt^2 + \delta_{ij} e^i_a e^j_b dx^a \wedge dx^b \,$ we determine the triads for our metric to be given by $\, e^i_a = a_i(t) \delta^i_a \,$, i.e.
\begin{equation}
e^1_1 = a_1(t) \quad , \quad e^2_2 = a_2(t) \quad , \quad e^3_3 = a_3(t)\,.
\label{eq:3.14}
\end{equation}

This allows us to calculate the spin connection $\Gamma^i_a$ which we need for the Ashtekar variables $A^i_a = \Gamma^i_a + \gamma K^i_a$ where $K^i_a$ is given by $K^i_a = e^b_i K_{ab} = h^{bc} e^i_c K_{ab}$. But we know that the spin connection vanishes for our flat spatial slice, so the three non-vanishing components of the Ashtekar variables are simply given by the (classical) relation
\begin{equation}
A^i_a = \gamma \, \dot{a}_i(t) \delta^i_a\,,
\label{eq:3.15}
\end{equation}

where $\delta^i_a$ is the co-triad compatible with Minkowski space. The densitized triads $E^a_i$ can be calculated via $\, E^a_i = \frac{1}{2} \varepsilon^{abc} \varepsilon_{ijk} e^j_b e^k_c \,$ with the result
\begin{equation}
E^1_1 = a_2(t) a_3(t) \quad , \quad E^2_2 = a_1(t) a_3(t) \quad , \quad E^3_3 = a_1(t) a_2(t)\,.
\label{eq:3.16}
\end{equation}

\subsection{Solution of the Bianchi I model}

In this subsection we present the Kasner solution for the model under consideration. We examine solutions to the equation
\begin{equation}
8 \pi G \, \sqrt{h} \, \mathcal{C} = a_1 \dot{a}_2 \dot{a}_3 + a_2 \dot{a}_1 \dot{a}_3 + a_3 \dot{a}_2 \dot{a}_1 = 0
\label{eq:3.17}
\end{equation}

as our initial values. Using the Einstein equations in their Hamiltonian form one can derive the time evolution of these initial values and by that construct the full spacetime metric. Equivalently, one can also try to solve the Einstein equations directly. We will use the second way by making the following ansatz
\begin{equation}
a_i(t) = t^{\kappa_i} \quad , \quad \dot{a}_i(t) = \kappa_i \, t^{\kappa_i - 1} = \frac{\kappa_i}{t} \, a_i(t) \quad , \quad \ddot{a}_i(t) = \kappa_i (\kappa_i - 1) \, t^{\kappa_i - 2} = \frac{\kappa_i (\kappa_i - 1)}{t^2} \, a_i(t)\,.
\label{eq:3.18}
\end{equation}

By solving $G_{\mu \nu} = 0$, i.e. plugging our ansatz into Eq.(\ref{eq:3.4}), we find the following equations for the exponents $\kappa_i$,
\begin{equation}
\begin{array}{l}
\kappa_1 \kappa_2 + \kappa_1 \kappa_3 + \kappa_2 \kappa_3 = 0\\
\left(\kappa_2 + \kappa_3\right)^2 - \kappa_2 \kappa_3 = \kappa_2 + \kappa_3\\
\left(\kappa_1 + \kappa_3\right)^2 - \kappa_1 \kappa_3 = \kappa_1 + \kappa_3\\
\left(\kappa_2 + \kappa_1\right)^2 - \kappa_2 \kappa_1 = \kappa_2 + \kappa_1\,,
    \end{array}
\label{eq:3.19}
\end{equation}

which have a non-empty real solution space. So we ask without loss of generality what are the functions $\kappa_2(\kappa_1)$ and $\kappa_3(\kappa_1)$ and the answer is
\begin{equation}
\kappa_2(\kappa_1) = \frac{1}{2} \left(1 - \kappa_1 \mp \sqrt{1 + 2 \kappa_1 - 3 \kappa^2_1}\right) \quad , \quad \kappa_3(\kappa_1) = \frac{1}{2} \left(1 - \kappa_1 \pm \sqrt{1 + 2 \kappa_1 - 3 \kappa^2_1}\right)\,.
\label{eq:3.20}
\end{equation}

Since we want the exponents to be real one finds that $\kappa_1 \in \left[-\frac{1}{3}, 1\right]$ and the most common choice for the Kasner exponents is $\kappa_1 = - \frac{1}{3}$ and $\kappa_2 = \kappa_3 = \frac{2}{3}$\,. However, note that we can choose different exponents than those ones. Furthermore, one finds that all the Kasner exponents satisfy
\begin{equation}
\kappa_1 + \kappa_2 + \kappa_3 = \kappa^2_1 + \kappa^2_2 + \kappa^2_3 = 1\,.
\label{eq:3.21}
\end{equation}

\section{Our Model}
\label{OurModel}

Our goal is to calculate the EPRL-FK/KKL spinfoam transition amplitude between two coherent states, which are defined on a simple graph and are peaked on a classical anisotropic geometry. We follow the proposal laid out in \cite{ourpaper}: Take a single spinfoam vertex to connect the two Daisy graphs, which carry the information about the geometry on two separate spatial slices. An important advantage of our graph is the fact that we consider only closed links, which leads to the cancellation of the gauge terms, and thus allows for an analysis of the information coming from the geometric term. i.e. the $\sl2c$ label, even for small spins, thus we are able to sum over \textit{all} spins. This summation leads to an `improved' normalization, as we use the same approximation in both the character of the group element and the amplitude calculation\footnote{If one cancels the gauge term the vertex amplitude just reduces to a normal coherent state. The character of the $\sl2c$ element which for large spin reduces to the originally found term $\exp(-izj)$, but differs for small spins. Taking the full character into account has the result that we don't find a decreasing transition amplitude for small volumes any more and thus can't interpret our result as leading to a singularity resolution any more neither in the anisotropic, nor in the isotropic case.}.

In contrast to the first paper of this series we now relax the requirement of isotropy by allowing for different coherent state labels for each link.
\begin{figure}[h!]
\includegraphics[width=0.4\textwidth]{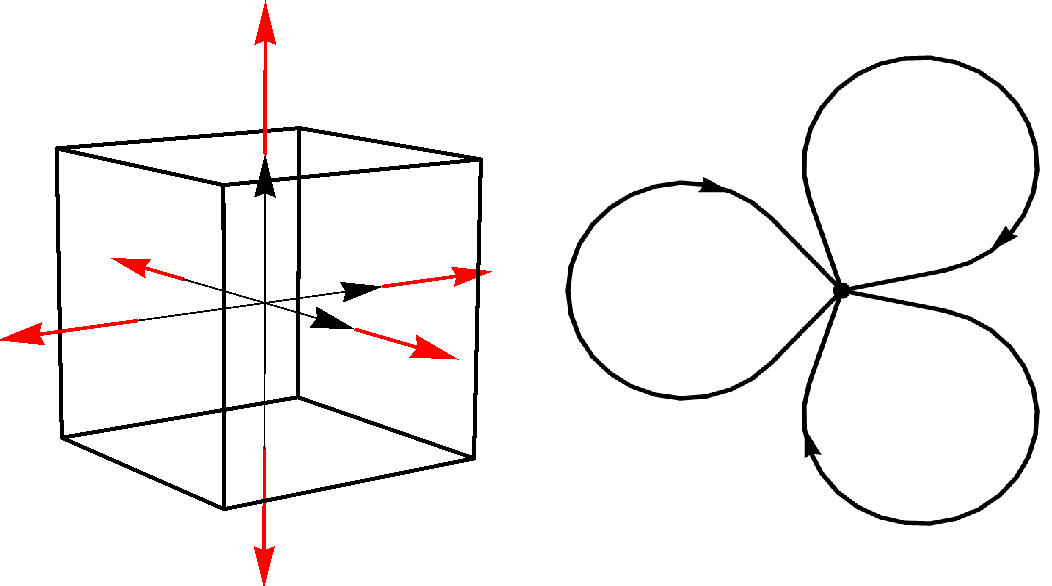}
\caption{\small Cube and Daisy graph}
\label{fig:1}
\end{figure}

We can specify the point in classical phase space where we want to peak our coherent state as follows. First note that the holonomy is unitary, $h_l \in SU(2)$ and the flux $E_l \in \mathfrak{su}(2)$. Remembering the left polar decomposition of $\sl2c$, i.e. $g = M u$, we can define a coherent state label $H_l$ per link by \cite{holo,coherentfrw}
\begin{equation}
H_l =  \exp \left(i \frac{E_l}{8 \pi G \hbar \gamma} t_l\right) h_l\left[A\right]\,.
\label{eq:4.1}
\end{equation}

The second decomposition of $H_l$ uses two $SU(2)$ elements $h_{\vec{n}_l}$ and $h_{\vec{n}^{\prime}_l}$ which, analogously to the $SU(2)$ elements of the Perelomov coherent states, correspond to the transformation of $\hat{e}_z$ into $\vec{n}_l$ and $\vec{n}^{\prime}_l$. Furthermore, a complex number $z_l$ is used whose real part is associated to the extrinsic curvature and its imaginary part is related to the area that is pierced by the link $l$ \cite{holo}
\begin{equation}
H_l = h_{\vec{n}_l} \, e^{-i z_l \frac{\sigma^3}{2}} \, h^{-1}_{\vec{n}^{\prime}_l}\,,
\label{eq:4.2}
\end{equation}

where we write $z_l= \R(z_l) + i \I(z_l)$ and $\sigma^3$ is the third Pauli matrix. We used Eq.(\ref{eq:4.2}) in \cite{ourpaper} to calculate the $\sl2c$ elements for our coherent states. In this paper we will calculate these elements using Eq.(\ref{eq:4.1}), cf. the comments following Eq.(\ref{eq:4.25}).

\subsection{Approximations}

For clarity of exposition, let us define the approximations we will make in performing the calculation. We follow the structure of \cite{Zakopane} in detailing these in terms of graph expansion, vertex expansion and large-spin limits. The nature of these limits has been discussed extensively in \cite{hellmann} to which we refer the reader for further discussion.

\subsubsection{Graph}

The graph we consider is the `Daisy graph', consisting of a single node which is both source and target to three links, \ref{fig:1}. This graph is dual to a cube whose opposing faces have been identified, and thus describes a compact spatial slice with three torus topology. In contrast to \cite{ourpaper} we shall allow independent spins on each link, thus breaking the isotropy of the model, yet retaining its homogeneity. This is the simplest anisotropic graph\footnote{Local rotational symmetry can be enforced as a further simplification by forcing equality between two of the three spins, however within GR this further restriction would yield only two possible solutions, and is therefore not very interesting from a cosmological viewpoint.}. 

\subsubsection{Vertex}

We work at the first order in vertex expansion, corresponding to a tree-level diagram in quantum field theory. The use of a single vertex is the subject of criticism in \cite{hellmann} though it remains a central approximation in \cite{Zakopane} and \cite{many, cosmoconstant}. Unlike usual field theories, the vertex does not carry a coupling constant and thus indication of what physical limit is being taken in truncating an expansion at this level. Hence, by one-vertex expansion we simply mean that our spinfoam consists of just one vertex. In \cite{sfexp1, sfexp2} a simple, discretized model is used to show that an expansion using a small number of vertices gives good agreement with the continuum dynamics both classically and in the quantum regime. In \cite{lewand} the sum of all graphs which contribute at the one vertex level was given for the dipole graph. However, fixing a boundary graph (motivated by a clear understanding of its dual representation) leaves just a single spinfoam history at this level. In this work we will take a position of agnosticism on this issue: We shall show that the first order in the vertex expansion is sufficient to reproduce the classical dynamics of our system, and leave an investigation of higher order corrections to future work. In \cite{hellmann} this approach was criticized further. It was stated that the factorization of the amplitude can be traced to the single vertex approximation and thus may prevent the evolution of degrees of freedom i.e. there will be no significant dynamics. Given the results of \cite{cosmoconstant} and this paper, both describing classically dynamical solutions, we believe that this criticism, at least within a certain semiclassical regime, is no longer justified. This can be compared with the findings of \cite{jacek}.

\subsubsection{Spins}

In previous works \cite{Zakopane,towards,many,lewand} the large spin approximation has played a significant role, with large spins being taken to mean semi-classical behaviour. However, as will be shown below, in our scenario this is unnecessary within the previously mentioned approximations, meaning, that we can calculate the full contribution without resorting to the large spin approximation. The reason is that due to the one-node structure of the Daisy graph the gauge part of the amplitude cancels and thus there is a reduction to simple coherent states for which one can calculate the heat kernel similar to the calculation of the norm of these states. However, as we will see, this comes with its own problems and its relevance for the general theory is unclear.

\subsection{Construction of the transition amplitude}

Now we assume that our spinfoam is made up from just one vertex. We can then simplify Eq.(\ref{eq:24}) by assuming that the holonomy $h_f$ is just a product of bulk holonomies $h_{vl}$ and link holonomies $h_l$ so that by integrating over the bulk holonomies they peak on the $h_l$. Thus we get
\begin{equation}
\bk{W}{\Psi} = \sum_{\sigma}{\int_{\su2}{dh_l \: A_v(h_l)} \, \Psi_{H_l}(h_l)} = \sum_{\sigma}{A_v(H_l)}\,,
\label{eq:4.3}
\end{equation}

where we have used for the second equality that the coherent state $\Psi_{H_l}(h_l)$ peaks $h_l$ on $H_l$. Following the procedure of \cite{towards} we restrict the calculation of Eq.(\ref{eq:4.3}) to a single spinfoam history. This step was further motivated in \cite{ourpaper} by our interest of the classical limit of Eq.(\ref{eq:24}) and thus we are looking, in some sense, for the spinfoam which corresponds to a classical trajectory. Then we can use Eq.(\ref{eq:22}) for the vertex amplitude in the holomorphic representation. By this we follow \cite{towards} but point out that it was argued in \cite{faceamplitude} that the face amplitude is given by $d_j = 2j+1$. Using now Eq.(\ref{eq:25}), i.e. we assume our boundary state to factorize as $\Psi = \ket{\Psi_{\text{out}}} \otimes \ket{\Psi_{\text{in}}}$, and also Eq.(\ref{eq:4.2}) we can write the transition amplitude as
\begin{equation}
W(\vec{z}_{\text{out}},\vec{z}_{\text{in}}) = \overline{A^{\text{out}}_v(H_l(z))} \, A^{\text{in}}_v(H_l(z))\,.
\label{eq:4.4}
\end{equation}

where in our case the number of links is 6 for the whole graph (two Daisy graphs) and $l=3$ for each single component.

\subsection{Holonomies and fluxes}

\subsubsection{Holonomies}
In order to calculate the holonomies and the fluxes, using the classical input provided in section \ref{Classical}, we consider our cube with one node at its center and with respective edge lengths $L_i$, $i \in \{1,2,3\}$. The metric of this 3d spatial slice is given by the induced metric $h_{ab}$ and the scale factors $a_i(t)$. Now, let us calculate the holonomy for the first link $l_1$
\begin{equation}
h_{l_1} \left[A\right] = \mathcal{P} \exp \left(\int_{l_1}{A}\right)\,.
\label{eq:4.5}
\end{equation}

For a path $e$, which is composed of two sub-paths $e = e_2 \circ e_1$, the holonomy has the following property:
\begin{equation}
h_e \left[A\right] = h_{e_2 \circ e_1} \left[A\right] = h_{e_2} \left[A\right] h_{e_1} \left[A\right]\,.
\label{eq:4.6}
\end{equation}

Thus, for our link $l_1$ we will calculate the holonomy for the half way from $\mathcal{O} = (0,0,0)$ to $P_1 = (\frac{L_1}{2},0,0)$ and then from $P_2 = (-\frac{L_1}{2},0,0)$ back to $\mathcal{O}$, (torodial topology or periodic boundary conditions). We say $l_1 = l^{\prime \prime}_1 \circ l^{\prime}_1$ and use the following parametrisations
\begin{equation}
l^{\prime}_1 : s \mapsto \begin{pmatrix}s\\0\\0\end{pmatrix} \quad , \quad l^{\prime \prime}_1 : s \mapsto \begin{pmatrix}-\frac{L_1}{2} + s\\0\\0\end{pmatrix}\,,
\label{eq:4.7}
\end{equation}

with $s \in \left[0,\frac{L_1}{2}\right]$. Now, the Ashtekar connection is given by $A = A^i_a \tau^i \text{d}x^a$ with $\tau^i = -\frac{i}{2} \sigma^i$, where $\sigma^i$ are the Pauli matrices. We have
\begin{equation}
\frac{dx(l^{\prime}_1(s))}{ds} ds = ds \quad , \quad \frac{dy(l^{\prime}_1(s))}{ds} ds = \frac{dz(l^{\prime}_1(s))}{ds} ds = 0
\label{eq:4.8}
\end{equation}

and thus we get (remember that the Ashtekar connection is diagonal for our model, Eq.(\ref{eq:3.15}), and thus the sum over the internal index is reduced to $i=1$)
\begin{equation}
h_{l^{\prime}_1} \left[A\right] = \mathcal{P} \exp \left(\int^{\frac{L_1}{2}}_0{A^1_1 \tau^1 \text{d}s}\right)\,.
\label{eq:4.9}
\end{equation}

Now, because the connection $A$ does not change along $l^{\prime}_1$, i.e. is independent of the spatial coordinates, we can drop the path-ordering symbol $\mathcal{P}$ which leads to
\begin{equation}
h_{l^{\prime}_1} \left[A\right] = \exp \left(A^1_1 \tau^1 \int^{\frac{L_1}{2}}_0{ \text{d}s}\right) =\exp \left(A^1_1 \tau^1 \frac{L_1}{2}\right)\label{eq:4.10}
\end{equation}

Using Eq.(\ref{eq:3.15}) for the components of the Ashtekar connection we find $A^1_1 = \gamma \dot{a}_1$. Thus, we get
\begin{equation}
h_{l^{\prime}_1} \left[A\right] = \exp \left(- i \, \frac{\gamma L_1 \dot{a}_1}{2} \, \frac{\sigma^1}{2}\right)
\label{eq:4.11}
\end{equation}

and analogously one finds for $l^{\prime \prime}_1$
\begin{equation}
h_{l^{\prime \prime}_1} \left[A\right] = \exp \left(- i \, \frac{\gamma L_1 \dot{a}_1}{2} \, \frac{\sigma^1}{2}\right)\,.
\label{eq:4.12}
\end{equation}

This gives us the following result for the holonomy along $l_1$
\begin{equation}
h_{l_1} \left[A\right] = h_{l^{\prime \prime}_1} \left[A\right] h_{l^{\prime}_1} \left[A\right] = \exp \left(- i \, \gamma L_1 \dot{a}_1 \, \frac{\sigma^1}{2}\right)\,.
\label{eq:4.13}
\end{equation}

With a similar procedure we get for the links $l_2$ and $l_3$
\begin{equation}
h_{l_2} \left[A\right]  = \exp \left(- i \, \gamma L_2 \dot{a}_2 \, \frac{\sigma^2}{2}\right) \qquad \text{and} \qquad h_{l_3} \left[A\right]  = \exp \left(- i \, \gamma L_3 \dot{a}_3 \, \frac{\sigma^3}{2}\right)\,.
\label{eq:4.14}
\end{equation}

Thus we find holonomies corresponding to the extrinsic curvatures of our faces\footnote{Note that our calculation of the holonomies (or the fluxes) does not resort to a toroidal topology. So at this point there is no difference whether we imagine a 6-valent node with six open links and periodic boundary conditions or a 6-valent node with three closed links. The reason why we consider toroidal topology is connected to the spinfoam vertex expansion. In this way we can easily forget about taking further nodes of a larger cubical lattice into account which would complicate the calculations. This problem is also related to the question of how to coarse grain over larger (boundary) graphs which carry the same information, as in a homogeneous setting for example. The first picture with open links seems to be closer to the canonical approach of \cite{quantred2, quantred}.}.

\subsubsection{Fluxes}
Let us now calculate the flux of the densitized triads, whose components we have calculated in Eq.(\ref{eq:3.16}), through the faces of our cube. We use Eq.(\ref{eq:3})
\begin{equation}
E(S_l) = \int_{S_l}{(\ast E)^j n^j}\,,
\label{eq:4.15}
\end{equation}

where $\ast$ denotes the Hodge dual, which converts our vector $E$ into a 2-form, ($\text{dim}(\Sigma)=3$), and $n^j = n^i \tau^i$ is a $\mathfrak{su}(2)$ valued scalar smearing function \cite{thiemann}. We have the following definitions
\begin{equation}
(\ast E) = (\ast E)^j \tau^j = (\ast E)^j_{a_1 a_2} \text{d}x^{a_1} \wedge \text{d}x^{a_2} \tau^j \quad , \quad (\ast E)^j_{a_1 a_2} = \varepsilon_{a a_1 a_2} E^a_j\,.
\label{eq:4.16}
\end{equation}

which allow us to calculate
\begin{equation}
(\ast E)^j = 2 E^3_j \, \text{d}x^1 \wedge \text{d}x^2 + 2 E^1_j \, \text{d}x^2 \wedge \text{d}x^3 + 2 E^2_j \, \text{d}x^3 \wedge \text{d}x^1\,.
\label{eq:4.17}
\end{equation}

By use of Eq.(\ref{eq:3.16}) for the components of the densitized triad we get for the first face $S_1$, which is dual to $l_1$
\begin{equation}
E_1(S_1) = \int_{S_1}{2 E^1_1 \tau^1 \, \text{d}x^2 \wedge \text{d}x^3} = 2 E^1_1 \tau^1 \int^{\frac{L_3}{2}}_{-\frac{L_3}{2}}{\int^{\frac{L_2}{2}}_{-\frac{L_2}{2}}{\text{d}y \text{d}z}} = -2 i E^1_1 L_2 L_3 \, \frac{\sigma^1}{2}\,.
\label{eq:4.18}
\end{equation}

Since also the triad is diagonal the only smearing function used for $S_1$ is $n^1 = \tau^1$ and the terms proportional to $E^2_2$ and $E^3_3$ give zero for $S_1$. Applying the same procedure to the other two links and using Eq.(\ref{eq:3.16}) we find the following results for the fluxes
\begin{equation}
E_1(S_1) = -2 i \, L_2 L_3 a_2 a_3 \, \frac{\sigma^1}{2} \quad , \quad E_2(S_2) = -2 i \, L_1 L_3 a_1 a_3 \, \frac{\sigma^2}{2} \quad , \quad E_3(S_3) = -2 i \, L_1 L_2 a_1 a_2 \, \frac{\sigma^3}{2}\,.
\label{eq:4.19}
\end{equation}

Thus our fluxes are proportional to the areas of the faces through which the edge passes. Making use of the polar decomposition, Eq.(\ref{eq:4.1}), we get our $SL(2,\mathbb{C})$ elements
\begin{equation}
H_{l}(z_l) = \exp \left(i \frac{E_l}{8 \pi G \hbar \gamma} t\right) \, h_{l} \left[A\right] = \exp \left(-i z_l \frac{\sigma^l}{2}\right)\,, \qquad (\text{no summation over }l)
\label{eq:4.20}
\end{equation}

where we have defined the complex number $z_l$ which is given for $l_1$ as
\begin{equation}
z_1 = \R(z_1) + i \I(z_1) = \gamma L_1 \dot{a}_1 + i \, \frac{L_2 L_3 a_2 a_3 t}{4 \pi G \hbar \gamma}
\label{eq:4.21}
\end{equation}

and the elements for $l_2$ and $l_3$ look analogously (permutation of $\{1,2,3\}$). Note that due to our graph we don't sum over $l$ in Eq.(\ref{eq:4.20}) and thus this model corresponds to a subspace of all admissible holonomies which is very similar to the procedure in the canonical quantum reduced model of \cite{quantred2, quantred}. Now we know how to evaluate Eq.(\ref{eq:4.20}) in the canonical basis $\ket{j,m}$ for all three links via an analytic continuation of the ordinary Wigner matrices where one can use for $l_1$ the following relation
\begin{equation}
e^{-i \frac{\pi}{2} \frac{\sigma^3}{2}} \, e^{i \alpha \frac{\sigma^1}{2}} \, e^{i \frac{\pi}{2} \frac{\sigma^3}{2}} = e^{i \alpha \frac{\sigma^2}{2}}\,.
\label{4.22}
\end{equation}

This allows us to determine for all three links the Wigner matrix elements. We notice that the coherent state labels obtained in the above section using Eq.(\ref{eq:4.1}) are slightly different from the labels we obtained in \cite{ourpaper} using Eq.(\ref{eq:4.2}). However, one finds that both labels lead to the same normalization. Also we find that both labels lead to the same result in the limit of large $\I(z)$. In the following section we will show which problem arises for these old labels for the links $l_1$ and $l_2$ as calculated by Eq.(\ref{eq:4.2}).

\subsection{Amplitude}
\label{amplitude}

Now let us collect everything together to calculate the transition amplitude Eq.(\ref{eq:4.4}) for our simple anisotropic model. Using the factorization of our amplitude we focus on just the incoming Daisy graph, which corresponds to $A^{\text{in}}_v(H_l(z))$, given by Eq.(\ref{eq:22}), with three links and the $\sl2c$ elements from Eq.(\ref{eq:4.20}). Then we have to normalize the amplitude as explained in Eq.(\ref{eq:26}). As shown in the appendix, Eq.(\ref{eq:031}), the norm Eq.(\ref{eq:16}) is equal for all three links and given by
\begin{equation}
\norm{\psi^t_{H_i}}^2_{n=0} = \frac{4 \sqrt{\pi} \, e^{t/4}}{t^{3/2}} \frac{\I(z_i)}{\sinh(\I(z_i))} \, e^{\frac{\I(z_i)^2}{t}}\,.
\label{eq:4.23}
\end{equation}

As has been previously noted, the fact that our graph has just one node and thus the source and target node for all three links are the same, leads to the simplification of the gauge contribution, i.e. $G_{s(l)} G^{-1}_{t(l)} = \mathbb{I}$, as $s(l)=t(l)$\footnote{Note that in our model $G_{s(l)} G^{-1}_{t(l)} = \mathbb{I}$ is a result of our special graph structure but that in more general situations $G_{s(l)} G^{-1}_{t(l)} \approx \mathbb{I}$ holds for small boost angles, as shown in \cite{jacek}.}. Following \cite{reg} we also have to drop one gauge integral so we are left with the following factor which characterizes our incoming geometry
\begin{equation}
W(z_1,z_2,z_3) = A^{\text{in}}_v(H_l(z_l)) = \prod^3_{l=1}{\sum_{2j \in \mathbb{N}_0}{(2j+1) \, e^{-\frac{t}{2} j (j+1)} \Tr(\mathcal{D}^{(j)}(H_l(z_l)))}}\,.
\label{eq:4.24}
\end{equation}

It is interesting to note here that with the approximations used so far (one-vertex spinfoam, just one spinfoam history, one node boundary graphs) the expression for the transition amplitude reduces to a product of two simple (gauge-variant) coherent states (or two delta functions), peaked at the $H_l$ and evaluated at the identity. This observation has the remarkable consequence that we do not require the large spin approximation which has been used in all the spinfoam cosmology investigations thus far. We can evaluate the character appearing in Eq.(\ref{eq:4.24}) explicitly using Eq.(\ref{eq:19}). For ease of comparison between this work and those prior, let us first make use of the large spin approximation, which means that we approximate the trace by $\exp(-i z_l j)$. This result can be easily reproduced from the full character, as we will show below. We will see that this is indeed a good approximation for the classical limit and the result is only changed for small spins. 

At this point we deviate from the $\sl2c$ labels used in \cite{ourpaper}, i.e. the ones calculated using Eq.(\ref{eq:4.2}), cf. Eq.(\ref{app:1}) and Eq.(\ref{app:2}). As we mentioned earlier, we recover the same normalization with those label and we can also apply the large spin approximation, as was done in \cite{ourpaper}, and get the same result. However, using Eq.(\ref{eq:19}) and Eq.(\ref{eq:20}) we find for the character of $H^{\text{old}}_1(z_1)$ and $H^{\text{old}}_2(z_2)$
\begin{equation}
\chi_j(H^{\text{old}}_1) = \chi_j(H^{\text{old}}_2) = \frac{i^{(2j+1)} - i^{-(2j+1)}}{i - (-i)} = \sin\left((2j+1) \, \pi /2\right)
\label{eq:4.25}
\end{equation}

because for those labels we have $\Tr(H^{\text{old}}_1) = \Tr(H^{\text{old}}_2) = 0$ which leads to $\lambda = i$. This means that despite the labels Eq.(\ref{app:1}) and Eq.(\ref{app:2}) initially depend on arbitrary $z_1$ and $z_2$ the character fixes them to $z_1 = z_2 = \pi$. Thus, the `old' labels are not equivalent to our new labels and somehow carry less information. In this sense we would consider the new labels as the correct physical ones for our model. Note again, that this difference doesn't play a role in the large spin, respectively the large $\I(z_1), \I(z_2)$ limit.  Furthermore we believe that the failure of the old labels is not a general problem of those states but merely a degeneracy, related to the fact that the scalar product between two Perelomov coherent states, which are peaked on anti-parallel normals, vanishes. On the level of the transition amplitude the factors corresponding to the first two links are suppressed when using the old labels. They reduce to non-zero numerical factors and thus do not matter for the amplitude. Henceforth we will thus focus on the improved labels calculated in Eq.(\ref{eq:4.20}) which take into account the entire spin contribution. We can now follow our explanation from \cite{ourpaper} how to approximate the trace for large $\I(z)$ or observe that for the labels from Eq.(\ref{eq:4.20}) we have 
\begin{equation}
H_1(z_1) = e^{-i z_1 \frac{\sigma^1}{2}} = \begin{pmatrix}\cos \left(\frac{z_1}{2}\right) & -i \sin \left(\frac{z_1}{2}\right)\\
 -i \sin \left(\frac{z_1}{2}\right) & \cos \left(\frac{z_1}{2}\right)\end{pmatrix}
\label{eq:4.26}
\end{equation}

\begin{equation}
H_2(z_2) = e^{-i z_2 \frac{\sigma^2}{2}} = \begin{pmatrix}\cos \left(\frac{z_2}{2}\right) & -\sin \left(\frac{z_2}{2}\right) \\
 \sin \left(\frac{z_2}{2}\right) & \cos \left(\frac{z_2}{2}\right)\end{pmatrix}
\label{eq:4.27}
\end{equation}

\begin{equation}
H_3(z_3) = e^{-i z_3 \frac{\sigma^3}{2}} = \begin{pmatrix}e^{-\frac{i z_3}{2}} & 0 \\
 0 & e^{\frac{i z_3}{2}}\end{pmatrix}
\label{eq:4.28}
\end{equation}

Now using Eq.(\ref{eq:19}) and Eq.(\ref{eq:20}) we find explicitly that all three links give the same contribution and we also find how to reproduce the result from the large $\I(z)$ approximation
\begin{align}
\chi_j(H_i) = \Tr(\mathcal{D}^{(j)}(H_i(z_i))) &= \frac{e^{i (2j+1) z_i/2} - e^{-i (2j+1) z_i/2}}{e^{i z_i/2} - e^{-i z_i/2}} = \frac{\sin \left((2j+1) \, z_i /2\right)}{\sin \left(z_i/2\right)}\notag\\[0.5\baselineskip]
&\approx \frac{- e^{-i (2j+1) z_i/2}}{ - e^{-i z_i/2}} = e^{-iz_ij} \quad , \quad \I(z_i) \gg 1
\label{eq:4.29}
\end{align}

Thus, we find for Eq.(\ref{eq:4.24}) the result
\begin{equation}
W(z_1,z_2,z_3) = \prod^3_{l=1}{\sum_{2j \in \mathbb{N}_0}{(2j+1) \, e^{-\frac{t}{2} j (j+1) - i z_l j}}}\,,
\label{eq:4.30}
\end{equation}

This is simplified through a gaussian approximation yielding the following result
\begin{equation}
W(z_1,z_2,z_3) = \prod^3_{l=1}{\frac{\sqrt{8 \pi}}{t^{3/2}} \: (-i z_l) \, \exp \left( \frac{(t/2 + i z_l)^2}{2t} \right)}\,.
\label{eq:4.31}
\end{equation}

Making use of the normalization presented in Eq.(\ref{eq:4.23}) allows us to calculate the normalized amplitude following Eq.(\ref{eq:26}).
\begin{equation}
A_v(H_l)_{\text{norm.}} = \frac{W(z_1,z_2,z_3)}{\norm{\psi^t_{H_1}}^2 \norm{\psi^t_{H_2}}^2 \norm{\psi^t_{H_3}}^2} = \frac{W(z_1) W(z_2) W(z_3)}{\norm{\psi^t_{H_1}}^2 \norm{\psi^t_{H_2}}^2 \norm{\psi^t_{H_3}}^2}
\label{eq:4.32}
\end{equation}

We can now focus on one factor to find
\begin{equation}
\frac{W(z_i)}{\norm{\psi^t_{H_i}}^2} = \frac{\frac{\sqrt{8 \pi} \, (p_i - i c_i)}{t^{3/2}} \: \exp \left( \frac{1}{2t} \left(\frac{t^2}{4} + i t (c_i + i p_i) - (c_i + i p_i)^2\right)\right)}{\frac{4 \sqrt{\pi} \, e^{t/4}}{t^{3/2}} \frac{p_i}{\sinh(p_i)} \, e^{\frac{p^2_i}{t}}}\,,
\label{eq:4.33}
\end{equation}

where we have defined $\, z_i = c_i + i p_i \,$. Now, since we used the large $\I(z_i)$ approximation in Eq.(\ref{eq:4.29}) we also have to approximate
\begin{equation}
\sinh(p_i) = \frac{e^{p_i} - e^{-p_i}}{2} \approx \frac{e^{p_i}}{2} \quad , \quad p_i \gg 1\,.
\label{eq:4.34}
\end{equation}

The application of the large spin limit in the calculation of the amplitude but not for the normalization, which is a feature of spinfoam cosmology work to date, leads to the dropping of the transition amplitude which we have interpreted as the singularity resolution \cite{ourpaper}. To apply the analysis consistently one should either use the large $\I(z)$ approximation for both character and within the $\sinh(p)$ or one uses the full $\sinh(p)$ and takes the full character into account, and then uses a numerical analysis instead of the gaussian approximation. Since the resolution of singularities within prior work was based upon taking the large spin approximation at the level of the transition amplitude but not the normalization, it is unclear whether such results will occur within the refined calculation in which the approximation is applied even-handedly or the full context\footnote{If one is willing, despite these arguments, to take the full $\sinh$-contribution into account and use the large $\I(z)$ approximation in the numerator one would find a decreasing behaviour of this amplitude for small scale factors which, along the lines of \cite{ourpaper}, could be interpreted as a singularity resolution also in the anisotropic case. Recent work \cite{CarloFrancescaRecent} however seems to suggest that singularity resolution might be a generic feature in spinfoam theory.}. Taking this into account we find for Eq.(\ref{eq:4.33})\footnote{Note that we have neglected a factor of $\left(1-i \frac{c_i}{p_i}\right)$ in front of Eq.(\ref{eq:4.35}) for the following reason. Using the scaling behaviour with respect to $\hbar$, following from Eq.(\ref{eq:4.21}), which does not depend on the specific holonomies and fluxes we are using, we find that $\frac{c_i}{p_i} \propto \hbar$ and thus vanishes for $\hbar \rightarrow 0$. Even if we include it in the action, it wouldn't change our final result since it would give a logarithmic contribution to the action.}
\begin{align}
\frac{W(z_i)}{\norm{\psi^t_{H_i}}^2} &= \frac{1}{\sqrt{8}} \, \exp \left(\frac{p_i}{2} - \frac{p^2_i}{2t} - \frac{c^2_i}{2t} - \frac{t}{8} + i \, \frac{c_i}{2} - i  \, \frac{c_i p_i}{t}\right)\notag\\
&= \frac{1}{\sqrt{8}} \, \exp \left(\frac{i}{\hbar} \, S_G[c_i,p_i]\right)\,,
\label{eq:4.35}
\end{align}

where we have defined
\begin{equation}
S_G[c_i,p_i] =  \hbar \, \frac{c_i}{2} - \hbar \, \frac{c_i p_i}{t} - i \hbar \left( \frac{p_i}{2} - \frac{p^2_i}{2t} - \frac{c^2_i}{2t} -  \frac{t}{8}\right)\,.
\label{eq:4.36}
\end{equation}

If we now use this for the normalized amplitude, Eq.(\ref{eq:4.32}) we find
\begin{equation}
A_v(H_l)_{\text{norm.}} = \frac{1}{8^{3/2}} \, \exp \left(\frac{i}{\hbar} \, S_G[c_1,p_1,c_2,p_2,c_3,p_3]\right)
\label{eq:4.37}
\end{equation}

where the real and the imaginary part of the action is given by
\begin{equation}
\R(S_G) = - \frac{\hbar}{t} \left(c_1 p_1 + c_2 p_2 + c_3 p_3\right) + \frac{\hbar}{2} \left(c_1 + c_2 + c_3\right)
\label{eq:4.38}
\end{equation}

\begin{equation}
\I(S_G) = - \, \frac{\hbar}{2} \left( p_1 + p_2 + p_3 - \frac{p^2_1 + p^2_2 + p^2_3}{t} - \frac{c^2_1 + c^2_2 + c^2_3}{t} -  \frac{3 t}{4}\right)
\label{eq:4.39}
\end{equation}

and if we reinsert the expression for the real and imaginary part of $z_i$ found in Eq.(\ref{eq:4.21}), i.e. $\,c_1 = \gamma L_1 \dot{a}_1 \,$ and $\, p_1 = \frac{L_2 L_3 a_2 a_3 t}{4 \pi G \hbar \gamma} \,$, we get\footnote{Note that in the deep quantum regime the classical relation Eq.(\ref{eq:3.15}) is altered in LQC. The Hubble rate is given $H_1 \equiv \frac{\dot{a}_1}{a_1} = \frac{1}{2 \lambda \gamma} \left(\sin(\bar{\mu}_1 c_1 - \bar{\mu}_2 c_2) + \sin(\bar{\mu}_2 c_2 + \bar{\mu}_3 c_3) + \sin(\bar{\mu}_1 c_1 - \bar{\mu}_3 c_3)\right)$ and the relation $c_i = \gamma L_1 \dot{a}_i$ is only recovered in the classical limit $\sin(x) \approx x$. The $\mu$'s are given by $\bar{\mu}_1 = \lambda / L_1 a_1$ and $\lambda \propto L_{P}$. Since we are looking for the classical limit of our model using Eq.(\ref{eq:3.15}) is justified.}
\begin{equation}
\R(S_G) = - \frac{L_1 L_2 L_3}{4 \pi G} \left(\dot{a}_1 a_2 a_3 + a_1 \dot{a}_2 a_3 + a_1 a_2 \dot{a}_3\right) + \frac{\hbar \gamma}{2} \left(L_1 \dot{a}_1 + L_2 \dot{a}_2 + L_3 \dot{a}_3\right)\,.
\label{eq:4.40}
\end{equation}

By writing $\text{Vol}(\Sigma_t) = \int_{\Sigma_t}{\sqrt{h} \, d^3y} = a_1 a_2 a_3 L_1 L_2 L_3$ for a cube with edge lengths $L_i$ and using Eq.(\ref{eq:3.9}) we can rewrite the first term of Eq.(\ref{eq:4.40}) using
\begin{equation}
a_1 a_2 a_3 L_1 L_2 L_3 \left(\frac{\dot{a}_1}{a_1} + \frac{\dot{a}_2}{a_2} + \frac{\dot{a}_3}{a_3}\right) = \int_{\Sigma_t}{\sqrt{h} \, K \, d^3y}\,,
\label{eq:new}
\end{equation}

which is the boundary term Eq.(\ref{eq:3.5}). Consider now Eq.(\ref{eq:4.4}) for the full amplitude between our initial and final state
\begin{equation}
W(\vec{z}_{\text{out}},\vec{z}_{\text{in}}) = \overline{A^{\text{out}}_v(H_l(z))} \, A^{\text{in}}_v(H_l(z))\,.
\label{eq:4.41}
\end{equation}

and note the complex conjugation of the first factor, \cite{oeckl1, oeckl2}. Using now Eq.(\ref{eq:4.37}) we find for the normalized amplitude
\begin{equation}
\mathcal{A}(H^{\text{out}}_l,H^{\text{in}}_l) = \frac{1}{8^3} \, \exp \left(\frac{i}{\hbar} \, S_G[z^{\text{out}}_l,z^{\text{in}}_l]\right)
\label{eq:4.42}
\end{equation}

with
\begin{align}
&S_G[z^{\text{out}}_l,z^{\text{in}}_l] = \frac{1}{4 \pi G} \int_{\Sigma_{\text{out}}}{\dot{a}_1 a_2 a_3 + a_1 \dot{a}_2 a_3 + a_1 a_2 \dot{a}_3 \: \, d^3y}\label{eq:4.43}\\[0.5\baselineskip]
&- \frac{1}{4 \pi G} \int_{\Sigma_{\text{in}}}{\dot{a}_1 a_2 a_3 + a_1 \dot{a}_2 a_3 + a_1 a_2 \dot{a}_3 \: \, d^3y} + \text{quantum corrections} + \text{imaginary part},\notag
\end{align}

which is the main result of this work. If we neglect the quantum corrections and the imaginary part for the moment and recall the classical action from section \ref{Classical}, Eq.(\ref{eq:3.23}),
\begin{equation}
S_G = - \frac{1}{8 \pi G} \int_{\mathcal{M}}{\left( a_1 \dot{a}_2 \dot{a}_3 + \dot{a}_1 a_2 \dot{a}_3 + \dot{a}_1 \dot{a}_2 a_3 \right) \: d^4x}\label{eq:4.44}
\end{equation}
\begin{equation}
+ \, \frac{1}{4 \pi G} \int_{\Sigma_{t_2}}{(\dot{a}_1 a_2 a_3 + a_1 \dot{a}_2 a_3 + a_1 a_2 \dot{a}_3) \, d^3y} \, - \, \frac{1}{4 \pi G} \int_{\Sigma_{t_1}}{(\dot{a}_1 a_2 a_3 + a_1 \dot{a}_2 a_3 + a_1 a_2 \dot{a}_3) \, d^3y}\,,\notag
\end{equation}

we find that we recover exactly the classical Hamilton function of the Bianchi I spacetime under consideration, which is exactly what to expect. This means, the action is evaluated on solutions of the field equations and thus the bulk term vanishes (for $\Lambda = \rho = 0$). What is left are the boundary terms for the initial and the final slice, i.e. the boundary data which select a specific classical trajectory. Now, what do we get if we insert the Kasner solution into the boundary terms? Using $\kappa_1 + \kappa_2 + \kappa_3 = 1$ one finds that the integrand is identically (for all $t > 0$) unity:
\begin{equation}
\dot{a}_1 a_2 a_3 + a_1 \dot{a}_2 a_3 + a_1 a_2 \dot{a}_3 = \kappa_1 t^{\kappa_1 - 1} t^{\kappa_2} t^{\kappa_3} + \kappa_2 t^{\kappa_1} t^{\kappa_2 - 1} t^{\kappa_3} + \kappa_3 t^{\kappa_1} t^{\kappa_2} t^{\kappa_3 - 1} = 1
\label{eq:4.45}
\end{equation}

and thus the (classical) Hamilton function vanishes for the Kasner solution. Since we know the solution of the equations of motion one can also give a relation between the initial and final scale factors and their derivatives:
\begin{equation}
a^{\text{out}}_i = \left(\frac{t_{\text{out}}}{t_{\text{in}}}\right)^{\kappa_i} \, a^{\text{in}}_i \qquad , \qquad \dot{a}^{\text{out}}_i = \left(\frac{t_{\text{out}}}{t_{\text{in}}}\right)^{\kappa_i - 1} \, \dot{a}^{\text{in}}_i
\label{eq:4.46}
\end{equation}

which does not conflict with Eq.(\ref{eq:4.45}) as one can easily show. More importantly though, this showes that to leading order the Hamilton function is (again) a sum of an initial and final contribution and thus the factorization of the amplitude can be justified in hindsight. 

Now, what about the quantum correction term coming from Eq.(\ref{eq:4.40})
\begin{equation}
-\frac{\hbar \gamma}{2} \left(L_1 \dot{a}_1 + L_2 \dot{a}_2 + L_3 \dot{a}_3\right)^{\text{out}} \, + \, \frac{\hbar \gamma}{2} \left(L_1 \dot{a}_1 + L_2 \dot{a}_2 + L_3 \dot{a}_3\right)^{\text{in}}\,?
\label{eq:4.47}
\end{equation}

Note first the proportionality with respect to the Barbero-Immirzi parameter. There are two interesting statements one can make, first that for vanishing $\gamma$, which would correspond to a non-quantized geometry, the quantum correction vanishes, and second, for imaginary $\gamma$ (self-dual Ashtekar connection) this part belongs to the imaginary part of the action and cannot be interpreted as a first quantum correction. Another argument for complex $\gamma$ was put forward in \cite{neiman2}. However, if we stick to real, non-zero $\gamma$ and overlook these unconventional cases the following questions arises: what is the modified (effective) dynamics that arises from this quantum corrected model? At this point it is tempting to make comparisons with LQC, the highly successful canonical counterpart to our theory. However, at this stage such a comparison would run into immediate difficulties. The primary problem is that our model does not consist of a complete expansion in some small parameter, such as $\hbar$, and therefore we do not know if further terms would dominate those so far obtained. Furthermore the additional boundary terms do not contribute to an effective bulk action, since they are total derivatives. Eq.(\ref{eq:4.45}) will be altered by the inclusion of this term, and therefore there will be modifications to the Kasner solution. If we model such modifications as $a_i'=a_i + \epsilon_i$ we find that satisfying the modified equation at all times would require $\epsilon_i$ to grow as an inverse power of $t$ as $t \rightarrow 0$ quickly breaking any perturbative scheme.  We therefore believe that these terms will not constitute the basis for comparison with the Bianchi I models of \cite{lqcbianchi1} (and references therein). Thus, we believe that the most promising path would be to really compare the probabilities for the transitions from one state to another as predicted by LQC and spinfoam cosmology which should be possible numerically. Furthermore, given the rather simple structure of our quantum correction we believe it's very important to understand how our results change after a refinement of the spinfoam 2-complex or avoiding factorizing amplitudes before trying to talk about a possible effective dynamics. Qualitatively, these problems seem to be related to the two limits in spinfoam theory, small curvature and low energy (related to large spins) as summarized in \cite{muxin}.

Now, lets get to the imaginary part. In a general context the appearance of an imaginary action might seem puzzling at first sight, however, it is by now a long known phenomenon in general relativity. It dates back to the calculations on Euclidean (Wick rotated) black holes \cite{gibbonshawking} and in the context of finite boundaries (in Lorentzian spacetimes) it is now well know that this term is generic. In \cite{neiman1} it was first discussed for null-bounded regions in the second order formalism and then generalized to arbitrary finit spacetime regions with arbitrary causal boundaries in the first order formalism, \cite{neiman2, neiman3}. The general result for the imaginary part of the on-shell action of the Einstein-Hilbert action plus York-Gibbons-Hawking boundary term for finite Lorentzian spacetime regions is \cite{neiman3}
\begin{equation}
\I(S) = \frac{1}{4} \sum_{\text{flips}}{\sigma_{\text{flips}}}\,,
\label{eq:4.48}
\end{equation}

where $\sigma_{\text{flips}}$ is given by $A/(4G)$ with $A$ being the flip-surface, i.e. the surface where the normal of the boundary changes from timelike to spacelike or vice versa. This results from a carful analysis of the York-Gibbons-Hawking-boundary term for finite regions and an analytic continuation of the boost angle, thus leading to a contribution of $-i \pi /2$ every time the boundary normal `flips'. In case one just deals with a finite region bounded by spacelike and timelike boundaries the flip surfaces reduce to (non-differentiable) corners. In this case one finds that
\begin{equation}
\I(S) = \frac{A_0}{8G}\,,
\label{eq:4.49}
\end{equation}

with $A_0$ being the area of the `null-corner'. Now, let us recall Eq.(\ref{eq:4.39}) 
\begin{equation}
\I(S_G) = - \, \frac{\hbar}{2} \left( p_1 + p_2 + p_3 - \frac{p^2_1 + p^2_2 + p^2_3}{t} - \frac{c^2_1 + c^2_2 + c^2_3}{t} -  \frac{3 t}{4}\right)\,.
\label{eq:4.50}
\end{equation}

We point out that the two quadratic terms result from our work with coherent states. This means, these are the terms which are responsible that the states are sharply peaked on the $c_i$ and $p_i$ variables\footnote{In prior work \cite{towards, cosmoconstant, ourpaper} the normalization was chosen such that both the linear and the quadratic terms in the $p_i$ variables do not appear. We think that our analysis of the imaginary term is in favour of the way we normalize the amplitude.}. Thus, we neglect them in this analysis\footnote{One finds that the term proportional to $p^2$ has dimension of area squared whereas the term proportional to $c^2$ has dimension of area$^{-1}$.}, together with the last term, which gives merely a phase. We are thus left with the term linear in the $p_i$ variables and by rewriting them in terms of the scale factor and edge lengths we find (factoring out the $1/G$ as in Eq.(\ref{eq:4.49})) indeed that the relevant term has dimension of an area
\begin{equation}
- \frac{\hbar}{2} \left( p_1 + p_2 + p_3 \right) = - \frac{t}{8 \pi G \gamma} \left( L_2 L_3 a_2 a_3 + L_1 L_3 a_1 a_3 + L_1 L_2 a_1 a_3 \right)\,.
\label{eq:4.51}
\end{equation}

These are of course very heuristic ideas and we leave a more detailed analysis for future research. For example in \cite{neiman2} it was pointed out, that only for $\gamma = \pm i$ one achieves a correspondence between the classical action for the 4-Simplex and the asymptotic result of the EPRL-FK model. But since we have neglected the quadratic terms and furthermore haven't really taken the timelike boundaries into account a real comparison between the two results is questionable. However, we think that a suitable generalization of our model has the potential to clarify some of these interesting questions.

\section{Discussion}
\label{discussion}
In this paper we analysed the dynamics of a homogeneous and anisotropic vacuum cosmology within the spinfoam formalism. We used coherent quantum states, defined on a simple graph adapted to the symmetries of a classical Bianchi I model, to calculate the EPRL-FK/KKL transition amplitude to first order in the vertex expansion and showed that the result is in accordance with the classically expected Hamilton function. Additionally we also find first quantum corrections and an imaginary term of the action which carries the right dimensions compared to the findings of \cite{neiman1, neiman3}. Overall, this is a crucial test of the spinfoam formalism in the cosmological context. It is entirely conceivable that the result could have shown an inconsistency, casting doubt on the foundations of the subject or the approximations used. Here we reproduce a non-trivial dynamical state by probing an anisotropic geometry with the Daisy graph. Through a detailed calculation of the transition amplitude we recover exactly the boundary term of the gravitational action on the initial and final slices. Note that the absence of matter means that at the classical level there should be no contribution from the bulk. Furthermore, the choice of a graph with a single node, which forms the source and target for all links, and the resultant cancellation of the $\sl2c$ group elements allowed us to investigate some of the questionable issues of prior work in spinfoam cosmology in more detail. First of all, we can explicitly evaluate the character and thus sum over the all spins\footnote{This summation becomes accessible numerically, in this work we still used a gaussian approximation to derive our results.}, removing the need for a large spin limit. This seems to support the idea that the large spin limit is not equivalent to \emph{the} classical limit in general but depends on the regime where it is applied and is thus merely a certain scaling limit. Thus we conclude that the large spin approximation is not the semiclassical approximation to dynamics. This indicates that the tree level graph gives classical behaviour, and (further) quantum corrections are to be found by looking at higher orders in the vertex expansion. This point was also mentioned in \cite{many}. Understandably, the cancellation of group elements and the subsequent reduction of the amplitude to a product of two coherent states comes with its own problems. Nonetheless, this simplification led us to the realization that it is necessary to choose a different normalization. With this normalization we are closer to ordinary QFT and also find a term in the complex action which has the correct dimension of an area and which would have not been found using the normalization method of prior papers. A key difference between the normalizations is that with the newer choice the evidence of singularity resolution is not present. In \cite{ourpaper} the normalization suppressed states with small volumes, which we interpreted as singularity resolution. However, with the improved normalization, these states reappear which indicates that this resolution was due to the normalization used, not underlying quantum dynamics, cf. also footnote 4.

Another main point of criticism about previous calculations in spinfoam cosmology was the factorization of the amplitude, \cite{hellmann}. Concerning this issue let us make the following remarks. First, the factorization is not necessarily a feature of the one vertex expansion but of not taking the face amplitude in Eq.(\ref{eq:24}) properly into account. This means that by considering larger boundary graphs one can even at the one vertex level recover the missing integration by a detailed analysis of the face amplitude $\delta(h_f)$ in Eq.(\ref{eq:24}). This will be of importance in follow up work. Secondly, since we are able to reproduce a classically dynamical spacetime, it seems that this factorization is not in conflict with such scenarios. However, our analysis at the classical level does not answer questions about the propagation of quantum degrees of freedom. We are explicitly working with coherent states, sharply peaked on boundary data of classical solutions, the bulk term vanishes identically (in vacuum) and thus the factorization can always be justified from the classical division into a initial and final boundary term, cf. Eq.(\ref{eq:3.23}). This problem also becomes apparent by noting that our amplitude reduces to a product of two coherent states. Comparing this with propagators in standard QFT it is clear that we are missing an integration. Now, the lesson to learn here is that this is a problem not just of our Daisy graph model but also the Dipole model and that for future investigations larger graphs, which allow for a proper inclusion of the face amplitude, and thus make up for the missing integration, are mandatory if one wants to take spinfoam cosmology seriously. It took the great simplicity of our model to realize this issue.

Additionally to these more `structural questions', such as the use of the correct normalization, a more detailed investigation of the imaginary part, the missing integration or the treatment of more complicated graphs and spinfoams, we believe that it is of particular importance to find ways of how to match such covariant quantum cosmological models with real world physics. This means, given a certain amplitude, how can one compare it to observations in our universe? In \cite{ourpaper} we have already mentioned that, given an initial state of the universe, one could consider a random walk in the space of scale factors with transition probabilities as described by the distribution obtained from the quantum amplitude. However, in order to make contact with the observable universe and the precise models of cosmology which fit observations well, spinfoam models must be made to include matter.

\section*{Acknowledgments}
DS gratefully acknowledges support from the John Templeton Foundation as part of the Establishing the Philosophy of Cosmology collaboration. JR thanks Carlo Rovelli, Eugenio Bianchi and especially Emanuele Alesci for useful discussions on the coherent state structure of our amplitude.

\begin{appendix}
\label{appendix}
\section{The old coherent state labels}
In \cite{ourpaper} we have calculated the labels of our coherent states using Eq.(\ref{eq:4.2}) and got the following results
\begin{equation}
H^{\text{old}}_1(z) = -i \left(\sin\left(\frac{z}{2}\right) \sigma^3 + \cos\left(\frac{z}{2}\right) \sigma^2\right) = \begin{pmatrix}
	- i \sin\left(\frac{z}{2}\right) & -\cos\left(\frac{z}{2}\right) \\
\cos\left(\frac{z}{2}\right) & i \sin\left(\frac{z}{2}\right)
\end{pmatrix}\,,
\label{app:1}
\end{equation}

\begin{equation}
H^{\text{old}}_2(z) = i \left(\cos\left(\frac{z}{2}\right) \sigma^1 - \sin\left(\frac{z}{2}\right) \sigma^3\right) = \begin{pmatrix}
	- i \sin\left(\frac{z}{2}\right) & i \cos\left(\frac{z}{2}\right) \\
i \cos\left(\frac{z}{2}\right) & i \sin\left(\frac{z}{2}\right)
\end{pmatrix}\,,
\label{app:2}
\end{equation}

\begin{equation}
H^{\text{old}}_3(z) = \begin{pmatrix}
	e^{-i \frac{z}{2}} & 0 \\
0 & e^{i \frac{z}{2}}
\end{pmatrix}\,.
\label{app:3}
\end{equation}

As explained in the main text one notices that for $l_1$ and $l_2$ these have vanishing trace which renders them not good labels if one takes the whole character into account. However, if one applies the large $\I(z)$ approximation these labels give the same result as the `good' labels which were calculated using Eq.(\ref{eq:4.1}).

\section{Calculation of the norm of the coherent states}

In this section we calculate the (gauge-variant) scalar product of two heat kernel coherent states following closely \cite{gcs2}. We will use the familiar coherent states given by Eq.(\ref{eq:11}) and if we consider two such states with labels $g, g' \in \sl2c$, each having a (left) polar decomposition as $g = Mu$ with $M$ hermitian and $u$ unitarty we can write their scalar product as
\begin{align}
\bk{\psi^t_g}{\psi^t_{g'}} &= \int_{\su2}{dh \, \overline{\psi^t_g(h)} \psi^t_{g'}(h)}\notag\\
&= \sum_{2j,2k \in \mathbb{N}_0}{(2j+1) (2k+1) \, e^{-\frac{t}{2} j (j+1)} e^{-\frac{t}{2} k (k+1)} \, \int_{\su2}{dh \, \overline{\Tr_j(gh^{-1})} \Tr_k(g'h^{-1})}}\,.
\label{eq:03}
\end{align}

One can now write the traces in the above equation in a basis, insert a unity between the coherent state labels and the argument and  use the unitarity and the orthogonality relation of the Wigner matrices to show that
\begin{equation}
\int_{\su2}{dh \, \overline{\Tr_j(gh^{-1})} \Tr_k(g'h^{-1})} \, = \, \frac{\delta_{jk}}{(2j+1)} \, \Tr_j(g^{\dagger}g')\,,
\label{eq:06}
\end{equation}

which gives for Eq.(\ref{eq:03})
\begin{equation}
\bk{\psi^t_g}{\psi^t_{g'}} = \sum_{2j \in \mathbb{N}_0}{(2j+1) \, e^{-t j (j+1)} \, \Tr_j(g^{\dagger}g')}\,.
\label{eq:07}
\end{equation}

If we use now $\, g^{\dagger} = u^{\dagger} M^{\dagger} = u^{-1} M \,$ and the cyclicity of the trace we get
\begin{equation}
\Tr_j(g^{\dagger}g') = \Tr_j(M M' h^{-1}) \qquad \text{with} \qquad h = u u'^{-1} 
\label{eq:08}
\end{equation}

from which the final expression
\begin{equation}
\bk{\psi^t_g}{\psi^t_{g'}} =  \psi^{2t}_{M M'}(h)
\label{eq:09}
\end{equation}

follows. From this we get for the norm (squared) of a heat kernel coherent state (for which $g=g'$)
\begin{equation}
\norm{\psi^t_g}^2 = \bk{\psi^t_g}{\psi^t_g} = \psi^{2t}_{M^2}(1)\,.
\label{eq:010}
\end{equation}

Let us now calculate the norm explicitly. We use Eq.(\ref{eq:07}) and also our knowledge that every character of $g \in \sl2c$ can be written as
\begin{equation}
\Tr_j(g) = \frac{\lambda^{(2j+1)} - \lambda^{-(2j+1)}}{\lambda - \lambda^{-1}}
\label{eq:011}
\end{equation}

with $\, \lambda = x + \sqrt{x^2 - 1} \,$ and $\, x = \frac{a+d}{2} \,$. Using $\, \lambda - \lambda^{-1} = \sqrt{x^2 - 1} \,$ we find
\begin{align}
\bk{\psi^t_g}{\psi^t_g} = \frac{1}{\sqrt{x^2 - 1}}\sum_{2j \in \mathbb{N}_0}{(2j+1) \, e^{-t j (j+1)} \, \left(\lambda^{(2j+1)} - \lambda^{-(2j+1)}\right)}\,.
\label{eq:012}
\end{align}

Now define $\, (2j+1) = n \,$, which allows us to write
\begin{equation}
\bk{\psi^t_g}{\psi^t_g} = \frac{e^{\frac{t}{4}}}{\sqrt{x^2 - 1}} \, \sum_{n \in \mathbb{N}}{n \, e^{-\frac{t}{4} n^2} \, \left(\lambda^n - \lambda^{-n}\right)} = \frac{e^{\frac{t}{4}}}{\sqrt{x^2 - 1}} \, \sum_{n \in \mathbb{Z}}{n \, e^{-\frac{t}{4} n^2} \, \lambda^n}\,.
\label{eq:013}
\end{equation}

The sum can now be calculated using the Poisson summation formula, which states that for a continuous Schwartz function $f$ with Fourier transform $\hat{f}$ the following holds.
\begin{equation}
\sum_{n \in \mathbb{Z}}{f(n)} = \sum_{k \in \mathbb{Z}}{\hat{f}(k)}\,.
\label{eq:014}
\end{equation}

Our function $\, f(x) = x \, e^{-\frac{t}{4} x^2} \, \lambda^x \,$ is obviously a Schwartz function, so we calculate its Fourier transform.
\begin{equation}
\hat{f}(k) = \int^{\infty}_{-\infty}{f(x) \, e^{-2 \pi i k x} \, dx} = e^{\frac{(\ln(\lambda) - 2 \pi i k)^2}{t}} \int^{\infty}_{-\infty}{x \, e^{-\left(\sqrt{\frac{t}{4}} x - \frac{(\ln(\lambda) - 2 \pi i k)}{\sqrt{t}}\right)^2} \, dx}
\label{eq:015}
\end{equation}

and by use of the following transformation
\begin{equation}
y = \sqrt{\frac{t}{4}} \, x - \frac{(\ln(\lambda) - 2 \pi i k)}{\sqrt{t}} \quad , \quad x = \sqrt{\frac{4}{t}} \left(y + \frac{(\ln(\lambda) - 2 \pi i k)}{\sqrt{t}}\right) \quad , \quad dx = \sqrt{\frac{4}{t}} \, dy
\label{eq:016}
\end{equation}

we find
\begin{equation}
\hat{f}(k) = \frac{4 \sqrt{\pi}}{t^{3/2}} \, \left(\ln(\lambda) - 2 \pi i k\right) \, e^{\frac{(\ln(\lambda) - 2 \pi i k)^2}{t}}\,.
\label{eq:017}
\end{equation}

Thus, the final result is
\begin{equation}
\bk{\psi^t_g}{\psi^t_g} = \frac{4 \sqrt{\pi}}{t^{3/2}} \frac{e^{\frac{t}{4}}}{\sqrt{x^2 - 1}} \, \sum_{k \in \mathbb{Z}}{\left(\ln(\lambda) - 2 \pi i k\right) \, e^{\frac{(\ln(\lambda) - 2 \pi i k)^2}{t}}} = \psi^{2t}_{M^2}(1)
\label{eq:018}
\end{equation}

and we can calculate the normalization of our new coherent state labels
\begin{equation}
g_1 = e^{-i z_1 \frac{\sigma^1}{2}} = \begin{pmatrix}\cos \left(\frac{z_1}{2}\right) & -i \sin \left(\frac{z_1}{2}\right)\\
 -i \sin \left(\frac{z_1}{2}\right) & \cos \left(\frac{z_1}{2}\right)\end{pmatrix} \, , \, g_2 = e^{-i z_2 \frac{\sigma^2}{2}} = \begin{pmatrix}\cos \left(\frac{z_2}{2}\right) & -\sin \left(\frac{z_2}{2}\right) \\ \sin \left(\frac{z_2}{2}\right) & \cos \left(\frac{z_2}{2}\right)\end{pmatrix}\,,
\label{eq:019}
\end{equation}

\begin{equation}
g_3 = e^{-i z_3 \frac{\sigma^3}{2}} = \diag \left( e^{-\frac{i z_3}{2}} , e^{\frac{i z_3}{2}} \right)\,.
\label{eq:021}
\end{equation}

So we first have to find their polar decomposition $\, g_i = M_i u_i \,$, which is simple because we have their explicit Wigner representation in terms of exponentiated Pauli matrices. We know how to calculate $M$ and $u$, namely
\begin{equation}
M_1 = \left(g_1 g^{\dagger}_1\right)^{\frac{1}{2}} = \left(e^{-i (z_1 - \bar{z}_1) \frac{\sigma^1}{2}}\right)^{\frac{1}{2}} = e^{\I(z_1) \frac{\sigma^1}{2}}\,,
\label{eq:022}
\end{equation}

\begin{equation}
u_1 = M^{-1}_1 g_1 = e^{- \I(z_1) \frac{\sigma^1}{2}} e^{- z_1 \frac{\sigma^1}{2}} = e^{- i \R(z_1) \frac{\sigma^1}{2}}
\label{eq:023}
\end{equation}

(same result for $g_2$ and $g_3$) and thus find the following matrices
\begin{equation}
M_1 = \begin{pmatrix}\cosh \left(\frac{\I(z_1)}{2}\right) & \sinh \left(\frac{\I(z_1)}{2}\right)\\
 \sinh \left(\frac{\I(z_1)}{2}\right) & \cosh \left(\frac{\I(z_1)}{2}\right)\end{pmatrix} \, , \, M_2 = \begin{pmatrix}\cosh \left(\frac{\I(z_2)}{2}\right) & -i \sinh \left(\frac{\I(z_2)}{2}\right)\\ i \sinh \left(\frac{\I(z_2)}{2}\right) & \cosh \left(\frac{\I(z_2)}{2}\right)\end{pmatrix}\,,
\label{eq:024}
\end{equation}

\begin{equation}
M_3 = \diag \left( e^{\frac{\I(z_3)}{2}} , e^{-\frac{\I(z_3)}{2}} \right)\,.
\label{eq:026}
\end{equation}

In order to calculate the norm, we have to square these matrices with the result, that you just replace $\, \frac{\I(z_i)}{2} \,$ by $\, \I(z_i) \,$. Then we conclude that for all three links we have the same
\begin{equation}
M^2 \qquad \Rightarrow \qquad x = \frac{a + d}{2} = \cosh \left(\I(z_i)\right) \qquad \text{and}
\label{eq:027}
\end{equation}

\begin{equation}
\lambda = x + \sqrt{x^2 - 1} = \cosh \left(\I(z_i)\right) + \sinh \left(\I(z_i)\right) = e^{\I(z_i)}
\label{eq:028}
\end{equation}

and hence
\begin{equation}
\sqrt{x^2 - 1} = \sinh \left(\I(z_i)\right) \qquad \text{and} \qquad \ln(\lambda) = \I(z_i)\,,
\label{eq:029}
\end{equation}

which leads to the following result for Eq.(\ref{eq:018})
\begin{equation}
\bk{\psi^t_{g_i}}{\psi^t_{g_i}} = \frac{4 \sqrt{\pi}}{t^{3/2}} \frac{e^{\frac{t}{4}}}{\sinh \left(\I(z_i)\right)} \, \sum_{k \in \mathbb{Z}}{\left(\I(z_i) - 2 \pi i k\right) \, e^{\frac{\left(\I(z_i) - 2 \pi i k\right)^2}{t}}}\,.
\label{eq:030}
\end{equation}

And if we just take the term $k=0$ in the sum we have
\begin{equation}
\bk{\psi^t_{g_i}}{\psi^t_{g_i}} = \frac{4 \sqrt{\pi} \, e^{\frac{t}{4}}}{t^{3/2}} \frac{\I(z_i)}{\sinh \left(\I(z_i)\right)} \, e^{\frac{\I(z_i)^2}{t}}\,,
\label{eq:031}
\end{equation}

which is the norm we have used in the main text. The restriction to $k=0$ is justified by the fact that the terms with $k \neq 0$ are exponentially supressed as can be seen upon rewriting the sum in Eq.(\ref{eq:030}) as
\begin{align}
\sum_{k \in \mathbb{Z}}{\left(p_i - 2 \pi i k\right) \, e^{\frac{\left(p_i - 2 \pi i k\right)^2}{t}}} &= p_i \, e^{\frac{p^2_i}{t}} + \sum_{k \in \mathbb{N}}{e^{\frac{p^2_i}{t}} \, \cos \left(\frac{4 \pi p_i k}{t}\right) \, e^{- \frac{4 \pi^2 k^2}{t}} \, \left(2 p_i + 4 \pi i k\right)}\notag\\[0.5\baselineskip] &= p_i \, e^{\frac{p^2_i}{t}} \left[1 + \sum_{k \in \mathbb{N}}{\cos \left(\frac{4 \pi p_i k}{t}\right) \, e^{- \frac{4 \pi^2 k^2}{t}} \, \left(2 + \frac{4 \pi i k}{p_i}\right)}\right]\,,
\label{eq:032}
\end{align}

where we have used $p_i = \I(z_i)$. The cosine is oscillatory and bounded by $\pm 1$ and the factor $\exp \left(- 4 \pi^2 k^2 / t \right)$ leads to the supression of the sum for all $k \neq 0$.

\end{appendix}

\bibliography{references}
\bibliographystyle{ieeetr}

\end{document}